\documentclass[aps,prd,reprint,twoside,nofootinbib,preprintnumbers,superscriptaddress]{revtex4-2}

\usepackage[T1]{fontenc}
\usepackage[utf8]{inputenc}
\usepackage{lmodern}
\usepackage{amsmath,amssymb,bm,mathtools} 
\allowdisplaybreaks[3]   

\usepackage{graphicx}
\graphicspath{{./}}       
\usepackage{microtype}    
\usepackage{balance}      
\usepackage[normalem]{ulem} 
\usepackage{tikz}
\usepackage{pgfplots}
\pgfplotsset{compat=1.17} 
\colorlet{RED}{red} 

\usepackage{hyperref}
\hypersetup{colorlinks=true,linkcolor=blue,citecolor=blue,urlcolor=blue}

\begin{document}
\title{A Quantum Gravitational Mechanism for Isotropization of de Sitter Cosmologies}

\author{Stephon Alexander}
\affiliation{Brown Center for Theoretical Physics and Innovation, Brown University, Providence, RI 02912, USA}

\author{Bruno Alexandre}
\affiliation{Brown Center for Theoretical Physics and Innovation, Brown University, Providence, RI 02912, USA}
\affiliation{Abdus Salam Centre for Theoretical Physics, Imperial College London, London SW7 2BZ, United Kingdom}

\author{Daine L. Danielson}
\affiliation{Center for Theoretical Physics, Massachusetts Institute of Technology, Cambridge, MA 02139, USA}
\affiliation{Black Hole Initiative \& Smithsonian Astrophysical Observatory, Harvard University, Cambridge, MA 02138, USA}

\author{David N. Spergel}
\affiliation{Flatiron Institute, New York, NY 10010, USA}
\date{\today}

\begin{abstract}
Today, the observable cosmos exhibits a remarkable degree of isotropy and plausibly began in a nearly isotropic initial state. The properties of the Lorentzian Chern-Simons-Kodama (CSK) functional can provide an understanding of this initial state. In gravity with a positive cosmological constant, the Chern-Simons-Kodama (CSK) wavefunctional is an exact, chiral solution of the quantum gravitational constraints. We suggest that the normalizability and other issues with this functional, if interpreted as a proper state of quantum gravity, instead suggest an embedding into a larger quantum gravitational completion, and recast the CSK functional as a gravitational sphaleron with observationally desirable properties.
By perturbing around the dominant de Sitter saddle of the wavefunctional with appropriate quantum gravitational boundary conditions, we find that for a closed universe the system is dynamically driven to spatial isotropy, while all anisotropic modes acquire positive quadratic curvature and are Gaussian-suppressed. The decay of this sphaleron therefore proceeds along an isotropic channel, providing an intrinsic quantum-gravitational mechanism for dynamical isotropization. This isotropization effect is robust under the inclusion of a slow-roll inflaton, and no analogous isotropic sphaleron exists for spatially flat or hyperbolic geometries. Taken together, these results recast the Lorentzian CSK functional as a chiral sphaleron that naturally prepares an approximately isotropic de Sitter background for inflation. Beyond this phenomenological study, we further suggest that the CSK functional can be understood as a boundary functional for a class of anomaly-free objects, including a complexified generalization of the Hartle-Hawking state.

\end{abstract}

\maketitle

\section{Introduction}

Cosmic inflation successfully explains why the universe appears smooth, nearly flat, and statistically simple, characterized by nearly Gaussian, homogeneous, and isotropic fluctuations, on large scales \cite{Guth}. Despite this success,
inflation does not by itself explain why the universe began in a sufficiently
smooth and nearly isotropic state for inflation to start.  Standard
inflationary models assume such initial conditions, and although inflation
can dilute some departures from isotropy, it does not generically eliminate
all possible anisotropies.  In particular, homogeneous but anisotropic
cosmological solutions show that directional structure can persist unless the
universe begins close to isotropy \cite{Wald1983,BarrowHervik2006,MaleknejadSheikhJabbari2012}
.

This limitation reflects how inflation is usually modeled: one treats the homogeneous background geometry semiclassically (often assuming an approximately FRW, low-shear patch) and quantizes the inflaton and the linearized scalar/tensor perturbations on that background. This framework robustly predicts the statistics of fluctuations given an inflationary background, but it does not by itself determine a measure---or even a preferred wavefunctional---over possible background geometries, nor explain why the universe began within the basin of attraction of an approximately isotropic state required for inflation to start.

In this work we explore a quantum gravitational mechanism by which certain anisotropic spacetimes are, in fact, dynamically\footnote{`Dynamics' are here defined relative to a natural choice of relational clock: the isotropic mode along which the universe expands. In intermediate steps we keep the ADM coordinate time $t$ (and lapse) explicit; when we use language such as ``evolution'', ``decay'', or ``rolling off'' we mean relational evolution obtained after deparametrizing with respect to a monotonic clock (in vacuum, the isotropic expansion/mean-curvature mode on the expanding branch, and in the slow-roll extension one may equivalently use~$\phi$).} disfavored relative to isotropic ones. Our analysis is based on a quantum
wavefunctional naturally associated with a universe dominated by a positive

cosmological constant supporting a chiral gravitational wavefunctional: the Chern--Simons--Kodama (CSK) functional.\footnote{The predicted degree of observable chiral asymmetry is sensitive to the completion into a full quantum-gravitational state, or a UV completion. Whether the universe today exhibits the requisite degree of chiral gravitational asymmetry for a given completion remains an open question, and a potential target for high-precision observations.} The CSK functional can be understood as part of a complexified generalization of the Hartle-Hawking no-boundary wavefunctional (see, e.g., Appendix \ref{app:plebanski-completion}), and in isotropic minisuperspace, the CSK functional is also related (via an appropriate transform and contour prescription) to the Vilenkin/tunneling cosmology \cite{Magueijo:2020ugp}. While the interpretation of the CSK functional as a proper quantum state may be viewed as problematical, the wavefunctional can nevertheless be probed to give physically sensible answers to questions that hold irrespective of how the functional is completed into a full state of quantum gravity. In this paper, therefore, we take a phenomenological approach towards the CSK functional, and apply it to questions of relevance to early universe cosmology.\footnote{In fact, it is plausible that the CSK functional may be more fully understood as a metastable configuration of spacetime fluctuations above a proper state in a theory of quantum gravity. One such scenario is discussed in Appendix \ref{app:plebanski-completion}, and we leave this possibility to ongoing and future analysis.} We {first develop the mechanism in homogeneous minisuperspace, where the full reduced Bianchi-type \cite{Bianchi1898trans2001, EllisMacCallum1969} phase space can be treated explicitly, and then extend the quadratic} analysis to {the physical transverse--traceless tensor modes about the closed de~Sitter saddle}.  Throughout, ``isotropization'' refers to {Gaussian} suppression of {anisotropic data} in the {appropriate contour-defined wavefunctional or outgoing flux}.  {The full nonperturbative contour completion remains} an important direction for future work, though one minimal example is given in Appendix~\ref{app:plebanski-completion}.

Our central result is that, for a closed universe with spherical spatial
geometry, the quantum dynamics possess a special saddle-point configuration. 
This configuration is unstable in exactly one direction, corresponding to an
overall isotropic degree of freedom, while all anisotropic distortions are
stable and probabilistically suppressed.  In this sense, the saddle acts as a gravitational analogue of a sphaleron: a metastable saddle separating adjacent large-gauge/Chern--Simons sectors, with outgoing (relational) evolution along a single unstable direction in configuration space. In the case of the gravitational CSK functional, as the isotropic mode decays away
from the sphaleron saddle, anisotropies are damped and the geometry is driven in the direction of
isotropic de~Sitter space. 

With respect to the minisuperspace Hamiltonian, this 
sphaleron-like configuration sits atop an energetic ``mountain pass'' \cite{Manton_1983, Klinkhamer_1984} through the complexified configuration space that connects distinct winding sectors of the $\mathrm{SO}(3)$ frame field (i.e., the triad) of the isotropic spatial slices, related by large {$\mathrm{SO}(3)$ gauge transformations}.\footnote{Note that while the principal $\mathrm{SO}(3)$-bundles over $S^3$ are classified by $\pi_2(\mathrm{SO}(3)) = 0$ so that every such bundle is trivial, the large-gauge group itself is not topologically trivial because {$\pi_3(\mathrm{SO}(3))=\mathbb{Z}$}.}
In the metric formulation, by contrast, no analogous noncontractible path exists because the relevant framing topology is quotiented out in metric configuration space~\cite{Ashtekar:1988sw}.\footnote{By enforcing an $\mathrm{SO}(3)$ gauge choice one can lift a path through the space of metrics to a corresponding family of triads. Even in that setting, however, a lift between distinct homotopy classes of global $\mathrm{SO}(3)$ framings would require the inclusion of degenerate ($\det g=0$) metric configurations, while degenerate metrics are typically not in the integration domain of the path integral~\cite{Knorr:2022config}.} In this sense, the self-dual connection formulation employed here is particularly natural for describing the sphaleron process and its decay towards isotropy.

When restricted to the anisotropic Bianchi~IX sector with
$S^{3}$ spatial topology, the Chern-Simons-Kodama functional is perturbatively normalizable about the
saddle-point configuration. The wavefunctional is well behaved in the
anisotropic directions and exhibits a controlled semiclassical profile peaked around the de Sitter neck,
supporting the isotropization mechanism identified here. 

Going beyond minisuperspace, a natural choice of boundary conditions for the remaining transverse-traceless graviton modes (given by a suitable completion of the integration contour through the complexified phase space) is sufficient to isotropize all remaining dynamical degrees of freedom of the metric at quadratic order in perturbation theory.

Taken together, this provides an intriguing mechanism for the emergence of isotropy
without assuming finely tuned initial conditions. This isotropization mechanism consistently embeds within a simple model of slow-roll inflation, in which a background scalar inflaton field drives a slowly varying \textit{effective} cosmological constant (which is therefore not constant, but rather evolves adiabatically).  By contrast, within the spatially flat (Bianchi~I) and hyperbolic (Bianchi~V) minisuperspace sectors considered here, no analogous
instability-driven isotropization is found in spatially flat or hyperbolic cosmologies,
highlighting the special role played by closed spatial in this isotropization process.

The remainder of the paper develops this picture in detail.  We analyze the
quantum dynamics of homogeneous but anisotropic cosmologies, identify the
unstable saddle that governs their evolution, and demonstrate explicitly how
anisotropies are suppressed.  We also discuss a physical interpretation of the
unstable mode in terms of metastable decay (hinting at the possibility of a ``resonance'' interpretation for the CSK functional), and comment on broader connections
to quantum gravity and early-universe cosmology.

\section{Canonical Quantization}
\label{sec:quantization}

In this section we outline the canonical quantization of general relativity in its self-dual formulation, following Ashtekar. We introduce the self-dual connection variables and show how they naturally lead to a polynomial Hamiltonian constraint. The non-self-adjoint nature of the connection operator will motivate the modified inner product later in Section \ref{sec:kodama}.

Our index conventions are as follows: Greek letters $\{\alpha,\beta,...\}$ denote spacetime indices and capital Latin letters $\{I,J,..\}$ denote internal Lorentz indices, both running from 0 to 3. Lowercase Latin letters from the beginning of the alphabet $\{a,b,...\}$ denote spatial world indices, and those from the middle of the alphabet $\{i,j,...\}$ denote spatial frame indices, both of which run from 1 to 3. We work in the metric signature $\{-,+,+,+\}$, and we define the Levi-Civita tensor in terms of the symbol as $\epsilon_{abc}=\sqrt{h}\varepsilon_{abc}$, where $h$ is the determinant of the spatial 3D metric $h_{ab}.$

Einstein's field equations can be derived from the action
\begin{equation}
S_{\rm EC} =\frac{1}{32\pi G} \int\epsilon_{IJKL} \, e^I\wedge e^J\wedge\left[ R^{KL} - \frac{\Lambda}{6} e^K\wedge e^L\right],
\end{equation}
where $e^I = e^{I}_{\alpha} \mathrm{d}x^\alpha$ is the tetrad one-form, $\Lambda$ is the cosmological constant, and the curvature two-form $R^{IJ}$ of the spin connection $\omega_{IJ}$ is given by
\begin{equation}
R^{IJ} = \mathrm{d}\omega^{IJ} + \omega^I_{\ K}\wedge \omega^{KJ}.
\end{equation}
Varying with respect to $\omega$ gives
\begin{equation}
D(\epsilon_{IJKL}e^K\wedge e^L) = 0,
\end{equation}
which implies that the torsion $T^I = \mathrm{d}e^I + \omega^I_{\ J}\wedge e^J$ vanishes, and hence the spin connection is determined by the tetrad and its derivatives. The equation of motion for the tetrad becomes
\begin{equation}
\epsilon_{IJKL}e^J\wedge \left(R^{KL} - \frac{\Lambda}{3}e^K\wedge e^L\right) = 0,
\end{equation}
which is equivalent to Einstein’s equations.

A complex self-dual reformulation is obtained by taking the connection valued in the self-dual algebra $\text{sl}(2,\mathbb{C})^+$,
\begin{equation}
A^i = i\omega^{0i} - \frac{1}{2}\epsilon^{ijk}\omega_{jk}.
\end{equation}
This is precisely the chiral self--dual Ashtekar connection:
\begin{equation}
A^i = \Gamma^i + i K^i,
\end{equation}
where
\begin{equation}
\Gamma^i \equiv -\tfrac{1}{2}\,\epsilon^{i}{}_{jk}\,\omega^{jk},
\qquad
K^i \equiv \omega^{0i}.
\end{equation}
Its complex conjugate would correspond to the anti-self-dual connection acting 
on right-handed spinors, so the CSK wavefunctional depends only on the self-dual (left-handed) chiral representation of the gravitational connection.
The associated curvature is
\begin{equation}
F^i = \mathrm{d}A^i + \frac{1}{2}\epsilon^{ijk}A_j\wedge A_k.
\end{equation}
The action in self-dual variables reads
\begin{equation}
S_{\rm SD} = \frac{i}{8\pi G}\int\left[\Sigma_i\wedge F^i - \frac{\Lambda}{6}\Sigma_i \wedge \Sigma^i\right],
\label{eqSDaction}
\end{equation}
where $\Sigma^i = i e^0 \wedge e^i - \frac{1}{2}\epsilon^{ijk} e_j \wedge e_k$. This action is classically equivalent to the Einstein-Cartan action but admits a simpler canonical structure.

\subsection{Hamiltonian Formulation}

We briefly review how the Ashtekar phase space variables arise naturally from the ADM Hamiltonian formulation of general relativity. The starting point is a $3{+}1$ decomposition of spacetime adapted to a foliation by spacelike Cauchy hypersurfaces.

Let $t$ be a smooth scalar function on spacetime such that each level set
$t=\mathrm{const}$ defines a spacelike Cauchy hypersurface $\Sigma_t$.
The gradient
\begin{equation}
n_\alpha \equiv -\,N\,\partial_\alpha t
\end{equation}
defines a future-directed unit timelike normal vector field to $\Sigma_t$, with
$N>0$ the lapse function ensuring the normalization
$g^{\alpha\beta}n_\alpha n_\beta=-1$.

Introduce coordinates $\{y^a\}$ on each hypersurface $\Sigma_t$.
The tangent vectors to $\Sigma_t$ are then
\begin{equation}
\hat e^\alpha{}_a
\equiv
\left(\frac{\partial x^\alpha}{\partial y^a}\right)_{t},
\end{equation}
which satisfy $n_\alpha \hat e^\alpha{}_a = 0$.
The vector field $t^\alpha$ generating time evolution between hypersurfaces
satisfies $t^\alpha \partial_\alpha t = 1$ and admits the standard ADM
decomposition
\begin{equation}
t^\alpha = N n^\alpha + N^a \hat e^\alpha{}_a ,
\end{equation}
where $N^a$ is the shift vector encoding the tangential displacement of
coordinates along $\Sigma_t$.

With this decomposition, the spacetime metric takes the familiar ADM form
\begin{equation}
\mathrm{d}s^2
=
- N^2 \mathrm{d}t^2
+
h_{ab}
\left(
\mathrm{d}y^a + N^a \mathrm{d}t
\right)
\left(
\mathrm{d}y^b + N^b \mathrm{d}t
\right),
\end{equation}
where
\begin{equation}
h_{ab}
\equiv
g_{\alpha\beta}\,\hat e^\alpha{}_a \hat e^\beta{}_b
\end{equation}
is the induced spatial metric on $\Sigma_t$.

To introduce triad variables, we write the spatial metric as
$h_{ab} = e_a^{\,i} e_b^{\,j}\delta_{ij}$,
where $e_a^{\,i}$ is a spatial triad.
In terms of the triad, the self-dual two-form is
\begin{equation}
\Sigma^{\,i}_{ab}
=
-\epsilon^{i}{}_{jk}\, e_a^{\,j}\, e_b^{\,k},
\end{equation}
which plays a central role in the Ashtekar formulation.

The canonical momentum conjugate to the triad is encoded in the densitized triad
$E^a{}_i = \frac{1}{2}\varepsilon^{abc}\epsilon_{ijk} e_b^{\,j} e_c^{\,k}$,
while the configuration variable is the self-dual Ashtekar connection
\begin{equation}
A^i{}_a = \Gamma^i{}_a[e] + i\,K^i{}_a ,
\end{equation}
where $\Gamma^i{}_a[e]$ is the torsion-free spin connection compatible with
$e_a^{\,i}$ and $K^i{}_a$ is the triad-projected extrinsic curvature.

we can define  the densitized triad
\begin{equation}
    E^a_i = -\frac{1}{2}\varepsilon^{abc}\Sigma_{ibc}=\sqrt{h}e^a_i.
    \label{eq:densitized_triad}
\end{equation}
The SD action (\ref{eqSDaction}) then becomes
\begin{widetext}
\begin{equation}
S_{\rm SD}
 = \frac{1}{8\pi G\, i}\!\int\! d^4x\,
 \Big[
   E^a_i\,\mathcal{L}_t A^i_a
   - \frac{iN}{2\sqrt{h}}\,
     \epsilon^{ijk}E^a_i E^b_j
     \!\left(F_{kab}
     + \frac{\Lambda}{3}\varepsilon_{abc}E^c_k\right)
   + (t^\alpha A^i_\alpha)\,\mathcal{D}_a E^a_i
   + N^a E^b_i F^i_{ba}
 \Big],
\label{eq:SSD}
\end{equation}
\end{widetext}
and one arrives at the important Poisson bracket
\begin{equation}
\{
A^i_a(\vec{x}),E^b_j(\vec{y})\} = 8\pi G i \, \delta^b_a \delta^i_j \delta^{(3)}(\vec{x}-\vec{y}).
\end{equation}
The resulting constraints are the Gauss constraint $\mathcal{D}_a E^a_i = 0$, the diffeomorphism constraint $E^b_i F^i_{ba}=0$, and the  Hamiltonian constraint
\begin{equation}
\mathcal{H} = \frac{1}{16\pi G\sqrt{h}}\epsilon^{ijk}E^a_i E^b_j
\left(F_{kab} + \frac{\Lambda}{3}\varepsilon_{abc}E^c_k\right)=0.
\end{equation}
This Hamiltonian constraint is polynomial in $A$ and $E$, a key advantage of the Ashtekar variables.

\section{Chern--Simons--Kodama Functional}
\label{sec:kodama}

In the connection formulation, $\hat{A}^i_a$ acts by multiplication and $\hat{E}^a_i$ acts as a functional derivative. Assuming $EEF$ operator ordering, the Hamiltonian acts as
\begin{equation}
\epsilon_{ijk} \frac{\delta}{\delta A_{ai}}\frac{\delta}{\delta A_{bj}}
\left(F^k_{ab} + \ell_{\rm Pl}^2 \frac{\Lambda}{3} 
\varepsilon_{abc}\frac{\delta}{\delta A_{ck}}\right)\Psi[A] = 0,
\end{equation}
which is the Wheeler de Witt equation, up to contact terms that vanish identically in the minisuperspace truncation.  
The exact solution for non-degenerate metrics is the \textit{Chern--Simons--Kodama} (CSK) state:
\begin{equation}
\Psi_{\mathrm {CSK}}[A] = N \exp\!\left(\frac{3}{\ell_{\rm Pl}^2\Lambda} S_{\rm CS}[A]\right),
\end{equation}
where
\begin{equation}
S_{\rm CS}[A] = \int {\operatorname{Tr}}\!\left(A\wedge dA + \frac{2}{3}A\wedge A\wedge A\right)
\end{equation}
is the Chern--Simons functional. {Here $A=A^iT_i$ is valued in the $SU(2)$ spin lift of the oriented $SO(3)$ frame bundle, and $\operatorname{Tr}$ denotes the fundamental $SU(2)$ matrix trace with $T_i=-i\sigma_i/2$, so $\operatorname{Tr}(T_iT_j)=-\delta_{ij}/2$.} This state is an exact solution of all quantum constraints in the self-dual connection formulation.

Interpreted as a proper 
state, the CSK functional has been controversial because the exponent in $\Psi_{\mathrm {CSK}}[A]$ is unbounded, implying non-normalizability with respect to the naive inner product
\begin{equation}
\langle \Psi | \Phi \rangle_{\text{naive}} = \int [dA] \, \overline{\Psi[A]}\Phi[A].
\end{equation}
However, this is not the appropriate physical inner product for recovering real-spacetime physics after passing to a complexified, self-dual representation of gravity.
The connection $A^i_a$ is complex, and must satisfy the \textit{reality conditions}
\begin{equation}
E^a_i = \bar{E}^a_i, \qquad A^i_a + \bar{A}^i_a = 2\Gamma^i_a(E),
\end{equation}
where $\Gamma^i_a(E)$ is the spin connection compatible with the triad. These conditions modify the measure on the space of connections.

To preserve these conditions at the quantum level, the physical inner product should incorporate a nontrivial measure functional (and/or an equivalent contour prescription) consistent with the reality conditions, as proposed in \cite{AF2023}:
\begin{equation}
\langle \Psi | \Phi \rangle =
\int [dA\, d\bar{A}]\, \overline{\Psi[A]} \, e^{-S(\Re A)} \, \Phi[A],
\end{equation}
with
\begin{equation}
e^{-S(\Re A)} \equiv
\int [dE]\, \exp\!\left[-\frac{1}{\ell_{\rm Pl}^2}\int e_i \wedge d_{\Re A}e^i\right],
\end{equation}
where $d_{\Re A} = d + [\Re{A}, \cdot]$ is the covariant derivative. 
With this choice, the measure is compatible with the reality conditions and (in Euclidean signature) can render the perturbative evaluation of $|\Psi_{\mathrm {CSK}}|^2$ well behaved; it is proportional to the torsion $T^i=d_{\Re A}e^i$.

Accordingly, questions of (non)normalizability should be assessed using an inner product/measure consistent with the reality conditions (such as the one above), rather than the naive one.
The appearance of the Chern--Simons functional in the
Kodama wavefunctional reveals a profound link between quantum
gravity and topological transitions. In gauge theory,
the Chern--Simons action classifies vacua (i.e.\ flat/pure-gauge configurations with $F=0$) by integer
winding number, with sphalerons mediating tunneling
between adjacent topological sectors. For generic configurations with $F\neq 0$, the Chern--Simons functional is continuous (well-defined only modulo $\mathbb{Z}$) and can take non-integer values.  An analogous
structure emerges in gravity: the Chern--Simons--Kodama
(CSK) functional can be interpreted as a quantum superposition
of framed geometries whose Chern--Simons invariants differ by
integer multiples of $8\pi^{2}$.  In the Euclidean
self--dual formulation, this correspondence is made
explicit through the existence of distinct
$\theta$--sectors,
\begin{equation}
\Psi_{\rm CSK}^{(\theta)}[A^{g}]
   = e^{i\theta w(g)}\,\Psi_{\rm CSK}^{(\theta)}[A],
\label{eq:theta_euclidean}
\end{equation}
where $w(g)\!\in\!\mathbb{Z}$ denotes the winding number of a large
$\mathrm{SU}(2)$ gauge transformation and
{$\theta = 24\pi^{2}/(\Lambda\ell_{\rm Pl}^{2})\bmod 2\pi$ in the normalization of Eq.~\eqref{eq:theta_euclidean} and the large-gauge shift used below}.
Each $\theta$ labels a distinct Euclidean topological vacuum,
analogous to the $\theta$--vacua of Yang--Mills theory, and the
CSK wavefunctional plays the role of a topologically protected
ground state within that sector.

While the Euclidean formulation provides a mathematically
well--defined realization of this topological structure, the
Chern--Simons functional is real and the inner product is
positive definite, the physically relevant case is the
Lorentzian one.  There, the Ashtekar connection
$A=\Gamma+iK$ becomes complex and the measure ceases to be
real, requiring a holomorphic inner product consistent with the
reality conditions. With this continuation, questions of the normalizability of the functional become subtle,
and it is possible that, at least for some values of $\Lambda$, the functional is better understood as a resonance than as a state. We leave this precise characterization for future work, instead focusing on the phenomenology of the CSK functional, for which this Lorentzian continuation
is essential: only in that setting do the $\theta$--sectors acquire
dynamical significance, linking quantum topological transitions
to semiclassical de~Sitter geometry and to possible
parity--violating observables in gravitational dynamics.
In this sense, the Kodama functional unites the algebraic structure
of topological $\theta$--vacua with the physical content of
Lorentzian quantum cosmology.
In the following section we
develop this correspondence further and show how
gravitational sphalerons emerge as finite--energy configurations
interpolating between flat self--dual configurations related by large-gauge transformations.

\section{Instantons, Sphalerons, and Lorentzian Unwinding}

Instantons are Euclidean tunneling solutions connecting topologically distinct vacua, while sphalerons are Lorentzian, unstable configurations sitting atop the potential barrier separating them.  In $SU(2)$ gauge theory, the Chern--Simons number
\begin{equation}
N_{\mathrm{CS}}
 = \frac{1}{8\pi^2}
\int_{S^3}{\operatorname{Tr}}\!\left(A\wedge dA+\tfrac{2}{3}A\wedge A\wedge A\right)
\label{eqCSnumber}
\end{equation}
labels the topological large-gauge sectors of the theory.  Neighboring vacua differ by integer values of $N_{\mathrm{CS}}$, and the sphaleron sits at $N_{\mathrm{CS}}=\tfrac12$.

\medskip
\noindent
Gibbons and Steif~\cite{GibbonsSteif} first identified a Lorentzian sphaleron in $SU(2)$ Yang--Mills theory on the Einstein static universe, showing that topological transitions can occur in real time through unstable, homogeneous gauge configurations at half-integer Chern--Simons number.  Their result motivates an analogous construction in gravity, where the self--dual Ashtekar connection plays the role of the complexified gauge field mediating Lorentzian ``unwinding'' between gravitational large-gauge/framing sectors.

\medskip
\noindent
An analogous structure appears naturally in a Lorentzian de~Sitter background,
\begin{equation}
ds^2=-N^2(t)dt^2+a^2(t)\,d\Omega_3^2,
\end{equation}
where $t$ is time, $a(t)$ is the scale factor, and $d\Omega_3$ denotes the spatial three--sphere metric.  For a homogeneous closed ($k=+1$) spatial section, the Ashtekar connection takes the form~\cite{AF}
\begin{equation}
A^i=(1+ib)\,\frac{\omega^i }{2}\equiv f(t)\,\omega^i,
\qquad 
b=\frac{\dot a}{N}.
\label{eq:Ashtekar_ansatz}
\end{equation}
This form is the most general homogeneous and isotropic $SU(2)$--invariant connection on $S^3$, consistent 
with the decomposition $A^i = \Gamma^i + iK^i$.  
The real part (equivalently the spin-connection piece $\Gamma^i=\omega^i/2$ for the closed $k=+1$ slicing) encodes intrinsic curvature while the imaginary part 
$ib$ represents the extrinsic curvature or the conformal Hubble rate.  
Although this ansatz reproduces the isotropic de~Sitter solution, it also 
admits straightforward generalizations to homogeneous but anisotropic geometries 
(such as diagonal Bianchi~IX), so it should be regarded as a \emph{homogeneous} minisuperspace reduction.

This ansatz defines the Lorentzian gravitational analogue of the Yang--Mills sphaleron: a homogeneous configuration whose real and imaginary components correspond to the intrinsic and extrinsic curvatures of de~Sitter space.  

Let us explicitly compute the Chern--Simons number for this homogeneous configuration.  On $S^3$ let
$\int_{S^3}\omega^1\wedge\omega^2\wedge\omega^3=16\pi^2$,
$d\omega^i=-\tfrac12\epsilon^{i}{}_{mn}\,\omega^m\wedge\omega^n$, and {the same $SU(2)$ spin-lift generators}
$T_i=-\tfrac{i}{2}\sigma_i$, with
${\operatorname{Tr}}(T_iT_j)=-\tfrac12\delta_{ij}$ and
${\operatorname{Tr}}(T_iT_jT_k)=-\tfrac14\epsilon_{ijk}$.
The Chern-Simons number is given by (\ref{eqCSnumber}) and substituting the Sphaleron ansatz yields,
\begin{align}
{\operatorname{Tr}}(A\wedge dA)
 &=\frac{f^2}{4}\,\epsilon_{ijk}\,\omega^i\wedge\omega^j\wedge\omega^k,\\
\frac{2}{3}{\operatorname{Tr}}(A\wedge A\wedge A)
 &=-\frac{f^3}{6}\,\epsilon_{ijk}\,\omega^i\wedge\omega^j\wedge\omega^k,
\end{align}
and hence
\begin{equation}
N_{\mathrm{CS}}(f)=3f^2-2f^3.
\end{equation}
Evaluating $N_{\mathrm{CS}}(f)$ at $f=0$, $\,\tfrac12$, and $1$ gives 
$0$, $\,\tfrac12$, and $1$, respectively, confirming that the midpoint 
$f=\tfrac12$ marks the sphaleron point at the top of the Chern--Simons 
potential barrier. Along this homogeneous one-parameter family, $N_{\mathrm{CS}}(f)$ is purely topological only at the flat endpoints $f=0,1$ (where $F=0$). For $k=+1$ these endpoints correspond to $b=\pm i$ and thus lie off the real Lorentzian section $b\in\mathbb{R}$ in the self-dual representation. In this sense the associated `vacuum-to-vacuum' sphaleron picture is naturally formulated on the complexified configuration space, even though the sphaleron midpoint $f=\tfrac12$ ($b=0$) lies on the real Lorentzian slice. The reality conditions are then implemented through the choice of physical inner product (equivalently, a contour/measure prescription) so as to extract the real Lorentzian geometric data relevant to physics.
Suggestions of a Gamow resonance interpretation for the associated unstable mode are discussed in Sec.~X.

\section{Hamiltonian Derivation of Sphaleron Dynamics}

Before turning to anisotropic configurations, it is useful to isolate the isotropic closed de Sitter sector and make explicit how the sphaleron picture emerges already in this minisuperspace. In this setting the Lorentzian Kodama functional reduces to a wavefunction of a single connection variable, whose local semiclassical saddle-point structure we extract by expanding to quadratic order about the closed de Sitter configuration. As we now show, this configuration sits at the top of an effective potential and therefore plays the role of a sphaleron rather than a genuine ground state. In this light, the non-normalizability of the Kodama functional in the isotropic sector might be better understood as a consequence of the instability of the Kodama functional about the self-dual saddle, suggesting a resonance interpretation. We leave the detailed investigation of this possiblity to future work.

The evolution of the sphaleron profile can be derived directly from the Hamilton equations of motion for the connection (in a fixed time gauge, with dynamics understood to arise relationally).  The Hamiltonian constraint for gravity with cosmological constant is
\begin{equation}
\mathcal{H}
 = \epsilon^{ijk}\,E^a_i E^b_j
   \!\left(
   F_{kab}
   + \frac{\Lambda}{3}\,\varepsilon_{abc}\,E^c_k 
   \right)
 = 0.
\label{eq:Ashtekar_H_constraint}
\end{equation}
Recall the general ansatz for the sphaleron is
{\begin{equation}
A^i_a=f(t)\,\omega^i_a,
\qquad 
E^a{}_i=a^2(t)\,\hat e^a_i\equiv p(t)\,\hat e^a_i,
\qquad p(t)=a^2(t),
\label{eq:homogeneous_ansatz}
\end{equation}}
where {$\hat e^a_i$ is} the {inverse triad of the unit round $S^3$ metric}.  {Thus $p(t)$} is {the isotropic densitized-triad amplitude, not} the spatial volume density{; in this normalization $p\propto a^2$, while $\sqrt h\propto a^3$.  The one-forms $\omega^i$ satisfy $d\omega^i+\tfrac12\epsilon^i{}_{jk}\omega^j\wedge\omega^k=0$}.
The curvature is then
\begin{equation}
F^i=\dot f\,dt\wedge\omega^i
+\tfrac12(f^2-f)\epsilon^i{}_{jk}\,\omega^j\wedge\omega^k.
\end{equation}

Substituting \eqref{eq:homogeneous_ansatz} into \eqref{eq:Ashtekar_H_constraint} gives{, up to the overall positive factor fixed by the lapse and volume normalization,} the reduced {isotropic} constraint
{\begin{equation}
\mathcal{H}_{\rm iso}\propto
 a\left[4\left(f-\frac{1}{2}\right)^2-1+\frac{\Lambda}{3}a^2\right]
 =a\left[-(1+b^2)+\frac{\Lambda}{3}a^2\right],
\label{eq:H_constraint_reduced}
\end{equation}
where in the last equality we used $f=(1+ib)/2$.  The precise normalization used below is obtained directly from the reduced action.}
Going back to the SD action (\ref{eq:SSD}), the other two constraints are trivially satisfied, and after integrating over $S^3$ one obtains the explicit form:
\begin{eqnarray}
    S_{\text{SD}}=\frac{3V_c}{8\pi G}\int dt\left[a^2\dot{b}-Na\left(-1-b^2+\frac{\Lambda}{3}a^2\right)\right],
\end{eqnarray}
where $V_c=2\pi ^2$. The full Hamiltonian is then given by
\begin{eqnarray}
    H=\frac{3V_c}{8\pi G}Na\left[-(1+b^2)+\frac{\Lambda}{3}a^2\right].
\end{eqnarray}
We can now see that the canonical conjugate variables are in fact $a^2$ and $b$, with the commutator
\begin{eqnarray}
    [b,a^2]=\frac{i\ell_P^2}{3V_c}.
\end{eqnarray}
For fixed $a$, the Hamiltonian is maximized in the isotropic extrinsic-curvature direction at $b=0$, corresponding to the Chern–Simons midpoint $f=1/2$. Imposing the constraint then selects $a=a_\star=\sqrt{3/\Lambda}$, the time-symmetric round $S^3$ neck of global de Sitter. Thus the configuration is sphaleron-like in the self-dual connection description: it possesses an outgoing isotropic channel, while the subsequent Bianchi~IX analysis will show that homogeneous anisotropic directions are quadratically confined. In the enlarged ADM phase space of the next section, this point reveals itself to be a turning point of the closed de Sitter trajectory rather than a static equilibrium of the full Hamiltonian constraint.

In the isotropic closed de Sitter sector, the configuration $b=0$ therefore represents a saddle point with a single unstable direction in minisuperspace. The Kodama functional associated with this configuration is naturally interpreted as a sphaleron-like saddle sitting at the top of the potential, rather than as a normalisable ground state. The decay along the negative mode corresponds to the universe “rolling off’’ this isotropic de Sitter sphaleron into classical Lorentzian FRW evolution, interpreted as relational evolution in the chosen clock.

\section{Bianchi IX}

Having established in the isotropic sector that closed de{~}Sitter space is a sphaleron configuration sitting at the top of an effective potential, we now embed this picture into the diagonal anisotropic Bianchi~IX minisuperspace model.  The {purpose of this section} is to {identify} the de{~}Sitter {neck inside the full homogeneous real phase space, not merely inside the connection variables.  This distinction} is {important because a genuine metric anisotropy changes both} the {triad} and the {spin connection.  We therefore keep} the {full diagonal Bianchi~IX variables} through {quadratic order and only then identify} the anisotropic {stability properties of the} saddle.

We start by considering a diagonal Bianchi{~}IX metric written in terms of the lapse function {$N(t)$}, three directional scale factors {$a_i(t)$}, and a basis of left-invariant one-forms {$\omega^i$ on $S^3$}:
{\begin{eqnarray}
    ds^2=-N^2(t)dt^2+\sum_{i=1}^3 a_i^2(t)(\omega^i)^2 .
\end{eqnarray}}
On this homogeneous background we introduce an {$\mathfrak{su}(2)_{\mathbb C}$-valued} Ashtekar connection aligned with the invariant one-forms{,
\begin{eqnarray}
    A^i=f_i(t)\omega^i, \qquad f_i=\frac{\Gamma_i+i b_i}{2} .
\end{eqnarray}
Here $b_i$ are the} real {extrinsic-curvature variables on the Lorentzian section, while $\Gamma_i$} is the {spin connection} compatible with the {diagonal} triad{.  For Bianchi~IX,
\begin{eqnarray}
    \Gamma_i=\frac{a_j}{a_k}+\frac{a_k}{a_j}-\frac{a_i^2}{a_ja_k},\qquad j\neq k\neq i .
\end{eqnarray}
In the isotropic limit $a_1=a_2=a_3$, one has $\Gamma_i=1$} and {hence $f_i=(1+i b_i)/2$, reproducing} the {closed FRW ansatz} of {Sec.~V.}

The curvature of the Ashtekar connection, {$F^i$}, splits into an electric piece proportional to {$\dot f_i$} and a magnetic piece quadratic in {$f_i$}:
{\begin{eqnarray}
    F^i=\dot f_i dt\wedge\omega^i+(f_jf_k-f_i)\omega^j\wedge\omega^k , \qquad j\neq k\neq i .
\end{eqnarray}
We introduce the diagonal triad variables
\begin{eqnarray}
    \Pi_i=a_j a_k, \qquad j\neq k\neq i ,
\end{eqnarray}
so that}, in {these homogeneous} variables, the Hamiltonian constraint takes {the} symmetric polynomial form
\begin{widetext}
{\begin{eqnarray}
\mathcal{H} &=& \Pi_1\Pi_2(f_1f_2-f_3)+\Pi_2\Pi_3(f_2f_3-f_1)+\Pi_1\Pi_3(f_1f_3-f_2)+\frac{\Lambda}{4}\Pi_1\Pi_2\Pi_3 .
\end{eqnarray}}
\end{widetext}

The critical {equations} with respect to the connection components {$f_i$} are
{\begin{eqnarray}
    \frac{\partial\mathcal{H}}{\partial f_i}=\Pi_i(\Pi_j f_j+\Pi_k f_k)-\Pi_j\Pi_k=0 .
\end{eqnarray}
It is useful to define $X_i\equiv \Pi_i f_i$.  The three equations then become
\begin{eqnarray}
X_j+X_k=\frac{\Pi_j\Pi_k}{\Pi_i} ,
\end{eqnarray}
and their solution is
\begin{eqnarray}
 f_i=\frac12\left(\frac{\Pi_k}{\Pi_j}+\frac{\Pi_j}{\Pi_k}-\frac{\Pi_j\Pi_k}{\Pi_i^2}\right) .
\end{eqnarray}
Using $\Pi_i=a_ja_k$, this is precisely
\begin{eqnarray}
    f_i=\frac{\Gamma_i}{2} .
\end{eqnarray}
Therefore the equations $\partial\mathcal{H}/\partial f_i=0$ do not} by {themselves select an isotropic triad.  They impose
\begin{eqnarray}
    b_i=0
\end{eqnarray}
for an arbitrary nondegenerate diagonal Bianchi~IX triad.  The isotropic de~Sitter neck is selected only after} the {triad dependence of the Hamiltonian constraint is included.

We now impose the Lorentzian reality conditions explicitly by substituting
\begin{eqnarray}
    f_i=\frac{\Gamma_i+i b_i}{2}
\end{eqnarray}
into the homogeneous constraint.  A useful check is that the spin-connection contribution to the symplectic one-form is exact in this truncation:
\begin{eqnarray}
    \sum_i \Pi_i\, d\Gamma_i=0 .
\end{eqnarray}
Thus the real variables $(\Pi_i,b_i)$ give the expected homogeneous ADM phase space, up to the common overall normalization.  The resulting real Hamiltonian constraint is
\begin{widetext}
\begin{eqnarray}
\mathcal{H}_{\rm ADM} &=& \frac14\left[\Lambda a_1^2a_2^2a_3^2+a_1^4+a_2^4+a_3^4-2a_1^2a_2^2-2a_1^2a_3^2-2a_2^2a_3^2\right] \nonumber\\
&&-\frac14\left[\Pi_1\Pi_2 b_1b_2+\Pi_1\Pi_3 b_1b_3+\Pi_2\Pi_3 b_2b_3\right] .
\end{eqnarray}
\end{widetext}
This expression is the diagonal Bianchi~IX ADM constraint written in the normalization inherited from the self-dual minisuperspace variables.

To expand around the round de~Sitter neck, introduce Misner anisotropy variables
\begin{eqnarray}
    a_i=a e^{\beta_i}, \qquad \beta_1+\beta_2+\beta_3=0,
\end{eqnarray}
with
\begin{eqnarray}
    \beta_1=\beta_+ +\sqrt3\,\beta_- ,&&\quad
    \beta_2=\beta_+ -\sqrt3\,\beta_- ,\\
    \beta_3&=&-2\beta_+ .
\end{eqnarray}
Then
\begin{eqnarray}
    \beta_1^2+\beta_2^2+\beta_3^2=6(\beta_+^2+\beta_-^2) .
\end{eqnarray}
A real metric anisotropy changes the spin connection already at} first {order:
\begin{eqnarray}
    \Gamma_i=1-3\beta_i+O(\beta^2), \qquad
    \delta f_i^{(\Gamma)}=-\frac32\beta_i .
\end{eqnarray}
This is} the {part} of {the homogeneous perturbation that is missed if one varies only the imaginary connection components.

For the extrinsic-curvature variables define the orthonormal combinations
\begin{equation}
\begin{aligned}
 c_0 &= \frac{1}{\sqrt3}(b_1+b_2+b_3),\\[4pt]
 c_1 &= \frac{1}{\sqrt2}(b_1-b_2),\\[4pt]
 c_2 &= \frac{1}{\sqrt6}(b_1+b_2-2b_3).
\end{aligned}
\end{equation}
Here $c_0$} is the isotropic {extrinsic-curvature mode}, {while $c_1$ and $c_2$ are} the {two diagonal shear modes.  Expanding} the {exact real constraint} to quadratic order in {$\beta_\pm$ and $c_i$ gives
\begin{eqnarray}
\mathcal{H}_{\rm ADM} &=& \frac34 a^4\left(-1+\frac{\Lambda}{3}a^2\right)-\frac14 a^4 c_0^2+\frac18 a^4(c_1^2+c_2^2) \nonumber\\
&&+6a^4(\beta_+^2+\beta_-^2)+O(3) .
\end{eqnarray}
This is the central minisuperspace stability result.  The isotropic extrinsic-curvature direction has the negative DeWitt sign, while both the shear momenta and the genuine metric anisotropies have positive quadratic coefficients.

Finally, the ADM interpretation becomes transparent in canonical Misner variables.  Up to an overall normalization, the real symplectic one-form may be written as
\begin{eqnarray}
    -\sum_i b_i\,d\Pi_i
    &=& -2\sqrt3\,a^2c_0\,d\alpha+\sqrt6\,a^2c_2\,d\beta_+\\&&+\sqrt6\,a^2c_1\,d\beta_- ,
\end{eqnarray}
where $\alpha\equiv\ln a$.  Thus we define
\begin{eqnarray}
    p_\alpha=-2\sqrt3\,a^2c_0,\quad
    p_+=\sqrt6\,a^2c_2,\quad
    p_-=\sqrt6\,a^2c_1 .
\end{eqnarray}
In these variables the quadratic constraint becomes
\begin{eqnarray}
\mathcal{H}_{\rm ADM} &=& \frac34 a^4\left(-1+\frac{\Lambda}{3}a^2\right)-\frac{p_\alpha^2}{48}+\frac{p_+^2+p_-^2}{48} \nonumber\\
&&+6a^4(\beta_+^2+\beta_-^2)+O(3) .
\label{eq:full_bianchi_ix_adm_quad}
\end{eqnarray}
At the de~Sitter neck,
\begin{eqnarray}
    a=a_\star=\sqrt{\frac{3}{\Lambda}},\qquad
    p_\alpha=p_+=p_-=\beta_+=\beta_-=0,
\end{eqnarray}
so the full homogeneous anisotropy block is
\begin{eqnarray}
\mathcal{H}_{\rm ani}^{(2)}=\frac{p_+^2+p_-^2}{48}+6a_\star^4(\beta_+^2+\beta_-^2).
\label{eq:anisotropy_positive_block}
\end{eqnarray}
It is positive definite.  Thus the suppression of anisotropy survives the inclusion of the triad and spin-connection perturbations required by the second-order metric description}.

{\section{Isotropic Instability and Anisotropic Confinement}\label{sec:confinement}
Equation~\eqref{eq:full_bianchi_ix_adm_quad} exhibits} the {saddle} structure of the homogeneous {real phase space}.  The isotropic {momentum $p_\alpha$} appears with {the negative sign, while the physical anisotropy variables $(\beta_\pm,p_\pm)$ have} a {positive-definite quadratic} Hamiltonian.  {Hence the de~Sitter neck has a single} unstable {homogeneous channel}, associated with {isotropic expansion/contraction, and a stable transverse Bianchi~IX anisotropy sector.

The anisotropic part of} the {constraint is
\begin{eqnarray}
    \mathcal{H}_{\rm ani}^{(2)}=\frac{p_+^2+p_-^2}{48}+6a_\star^4(\beta_+^2+\beta_-^2),
\end{eqnarray}
which is the Hamiltonian of two stable homogeneous anisotropy oscillators.  This is the ADM form of the isotropization mechanism: homogeneous shear and real metric anisotropy are both energetically confined}, while the only {unstable} direction is the isotropic one{.

\begin{figure}[t]
\centering
\begin{tikzpicture}
\begin{axis}[
    width=0.9\linewidth,
    view={35}{30},
    xlabel={$p_{\alpha}$ (isotropic momentum)},
    ylabel={$p_{+}$ (anisotropic momentum)},
    zlabel={$\mathcal{H}^{(2)}$},
    colormap/viridis,
    domain=-2:2,
    domain y=-2:2,
    samples=45,
    samples y=45,
]
\addplot3[
    surf,
    shader=interp,
]
{ -3*x^2 + 2*y^2 };
\end{axis}
\end{tikzpicture}

\caption{Saddle structure of the quadratic homogeneous Hamiltonian in an isotropic momentum direction and one anisotropic momentum direction.  The isotropic direction has the negative DeWitt sign, while the anisotropic directions are stabilized.  The metric anisotropy variables $\beta_\pm$, not shown in this two-dimensional visualization, also appear with positive quadratic coefficients in Eq.~\eqref{eq:anisotropy_positive_block}.}
\label{fig:sphaleron_saddle}
\end{figure}
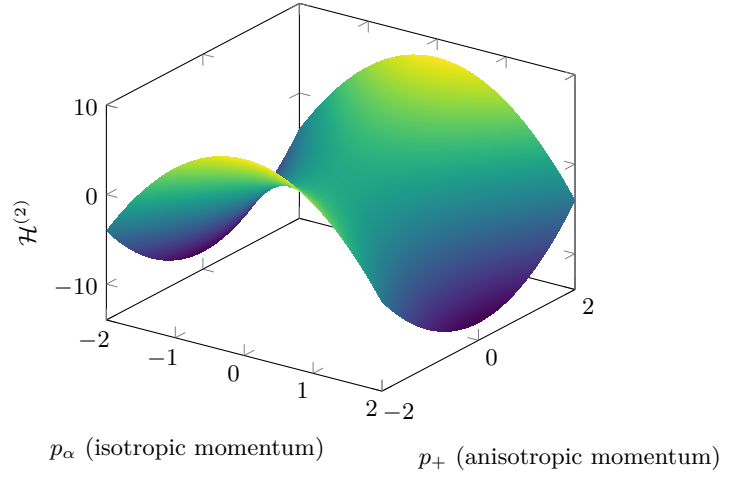

This structure is sphaleron-like} in the {framed self-dual connection variables: the configuration lies at the midpoint of the Chern--Simons path between adjacent large-gauge/framing sectors}.  In { ADM metric variables it can be described as the time-symmetric round $S^3$ neck of global de~Sitter, with a negative isotropic direction and positive transverse homogeneous anisotropy directions.  The outgoing relational evolution away from} this {neck is therefore naturally funneled along the isotropic channel}, {while homogeneous anisotropies are suppressed by} the {positive quadratic form.  This is the minisuperspace analogue} of {a gravitational sphaleron decay.

This ADM stability result, in turn, generates a probabilistic suppression of anisotropies in the Chern--Simons--Kodama wavefunctional.  {The} Chern--Simons functional on {diagonal Bianchi~IX data may be written, up to} the {same overall normalization used above}, {as
\begin{eqnarray}
    \mathcal{N}_{\rm CS}(f_1,f_2,f_3)=4\left(f_1^2+f_2^2+f_3^2-2f_1f_2f_3\right).
\end{eqnarray}
At} the {isotropic point $f_i=1/2$}, one {has $\mathcal{N}_{\rm CS}=2$.  If one varies only} the {imaginary connection components, $f_i=1/2+i b_i/2$, then
\begin{eqnarray}
    \mathcal{N}_{\rm CS}=2+i\sqrt3\,c_0-\frac32(c_1^2+c_2^2)+O(3),
\end{eqnarray}
and the connection-polarization CSK functional is Gaussian in the shear variables $c_1,c_2$:
\begin{widetext}
\begin{eqnarray}
    \Psi_{\mathrm {CSK}}[c_i] = \mathcal{N}\exp\left[\frac{3V_c}{\ell_{\rm Pl}^2\Lambda}\left(2+i\sqrt3\,c_0-\frac32(c_1^2+c_2^2)+O(3)\right)\right].
\end{eqnarray}
\end{widetext}
This reproduces the shear suppression found in the connection variables}.  {However, a real metric anisotropy also perturbs} the {spin connection}.  {Substituting} the full {real section
\begin{eqnarray}
    f_i=\frac{\Gamma_i(\beta)+i b_i}{2}
\end{eqnarray}
into the} Chern--Simons functional {gives
\begin{eqnarray}
\mathcal{N}_{\rm CS}&=&2+i\sqrt3\,c_0+72(\beta_+^2+\beta_-^2)-\frac32(c_1^2+c_2^2) \nonumber\\
&&-9i\sqrt6\left(\beta_-c_1+\beta_+c_2\right)+O(3) .
\label{eq:bare_csk_real_section_bianchi}
\end{eqnarray}
Thus the bare connection-space modulus evaluated directly on $A=\Gamma(E)+iK$ }, being complex, is not yet the physical probability density on real metric anisotropies.  The physical metric-polarization weight is obtained only after imposing the reality conditions through the connection-to-triad transform or the equivalent contour/measure.

{In the finite-dimensional Bianchi~IX truncation this transform takes the schematic form
\begin{eqnarray}
    \widetilde\Psi(\Pi)=\int_{\Gamma_\Pi}\prod_i df_i\,
    \exp\left[\frac{3V_c}{\ell_{\rm Pl}^2\Lambda}\left(\mathcal{N}_{\rm CS}(f)-\frac{2\Lambda}{3}\sum_i \Pi_i f_i\right)\right].
\end{eqnarray}
At the de~Sitter neck $\Pi_i=3/\Lambda$, the saddle is $f_i=1/2$.  For anisotropic real boundary triads, $\Pi_i=(3/\Lambda)e^{-\beta_i}$, the exponent on the Lorentzian real section expands as
\begin{eqnarray}
\Phi(\beta,c)&=&-1+60(\beta_+^2+\beta_-^2)-\frac32(c_1^2+c_2^2) \nonumber\\
&&-8i\sqrt6\left(\beta_-c_1+\beta_+c_2\right)+O(3),
\end{eqnarray}
where $c_0$ and the isotropic direction have been suppressed.  The Gaussian integral over the real shear variables $c_1,c_2$ is convergent and gives
\begin{eqnarray}
\int dc_1dc_2\,e^{\frac{3V_c}{\ell_{\rm Pl}^2\Lambda}\Phi(\beta,c)}
\propto
\exp\left[-\frac{12V_c}{\ell_{\rm Pl}^2\Lambda}(\beta_+^2+\beta_-^2)\right].
\end{eqnarray}
Consequently,
\begin{eqnarray}
\frac{|\widetilde\Psi(\beta_+,\beta_-)|^2}{|\widetilde\Psi(0,0)|^2}
=\exp\left[-\frac{24V_c}{\ell_{\rm Pl}^2\Lambda}(\beta_+^2+\beta_-^2)+O(\beta^3)\right].
\label{eq:metric_bianchi_gaussian}
\end{eqnarray}
The homogeneous metric anisotropies are therefore Gaussian-suppressed after the Lorentzian reality conditions are implemented in metric polarization.

\begin{figure}[t]
    \centering
    \IfFileExists{bowl2.png}{%
    \includegraphics[width=0.45\textwidth]{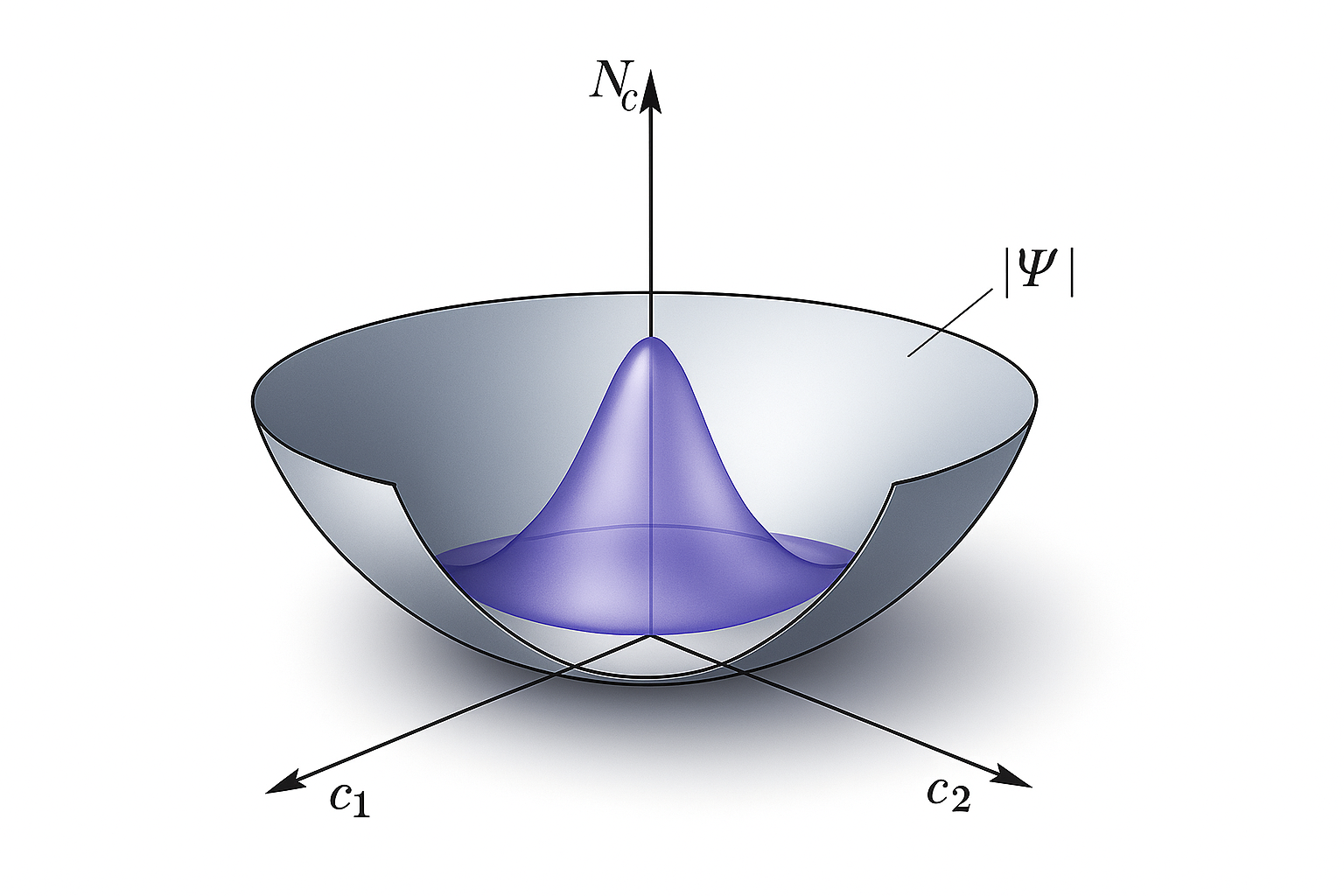}%
    }{%
    \fbox{\texttt{bowl2.png} not found}%
    }
    \caption{Connection-polarization visualization of the Chern--Simons functional in the shear-anisotropy plane spanned by $(c_1,c_2)$.  The bare CSK functional is Gaussian in the shear directions.  When the spin-connection perturbations associated with real metric anisotropies are also included, the physical metric-polarization transform gives the Gaussian suppression in $(\beta_+,\beta_-)$ shown in Eq.~\eqref{eq:metric_bianchi_gaussian}.}
    \label{fig:bowl2}
\end{figure}

\section{Isotropization Beyond Minisuperspace}
\label{sec:beyond-minisuperspace}

The preceding sections establish the isotropization mechanism in the full reduced homogeneous phase space.  We now ask how much of this structure survives when the homogeneous truncation is relaxed.  {To answer this, we carry the quadratic analysis one step beyond minisuperspace and identify the role played by the path integral contour in generating isotropization.}  The result is instructive.  The bare chiral CSK functional, evaluated with the Lorentzian reality conditions on the naive real contour, does not by itself suppress every transverse tensor mode: {with the tensor-curl convention used below, the sector $s<0$ is damped while the sector $s>0$ is anti-damped.}  This intermediate obstruction is the perturbative manifestation of the familiar fact that the bare holomorphic CSK functional should not, by itself, be interpreted as a complete physical state.  With a transverse no-boundary contour, however, the signed curl kernel is replaced by its positive spectral branch, and the suppression of anisotropy becomes completely general at this order.

We work at the round de~Sitter neck,
\begin{eqnarray}
    h_{ab}^{(0)}=a_\star^2\gamma_{ab},\qquad
    K_{ab}^{(0)}=0,
    \qquad
    a_\star=\sqrt{\frac{3}{\Lambda}},
\end{eqnarray}
where $\gamma_{ab}$ is the unit round metric on $S^3$.  A general perturbation of the real Lorentzian data can be written as
\begin{eqnarray}
    h_{ab}=a_\star^2(\gamma_{ab}+q_{ab}),
    \qquad
    K_{ab}=\delta K_{ab}.
\end{eqnarray}
After imposing the linearized momentum and Hamiltonian constraints and quotienting spatial diffeomorphisms, the physical inhomogeneous pure-gravity modes are transverse-traceless tensors,
\begin{eqnarray}
    D^a q_{ab}=0,
    \qquad
    \gamma^{ab}q_{ab}=0.
\end{eqnarray}
It is useful to diagonalize the tensor-curl operator
\begin{eqnarray}
{
    ({\cal C}T)_{ab}=\epsilon_{(a}{}^{cd}D_cT_{b)d}
}
\end{eqnarray}
on this transverse-traceless (TT) subspace.  We choose normalized eigenmodes $T^{(s)}_{ab}$ satisfying
\begin{eqnarray}
    {\cal C}T^{(s)}_{ab}=sT^{(s)}_{ab},
    \qquad
    s=\lambda\rho,
    \qquad
    \lambda=\pm,
    \qquad
    \rho>0.
\end{eqnarray}
For the round $S^3$ one may label $\rho$ by the usual tensor harmonic level, but only the sign structure will be important here.  Expanding
\begin{eqnarray}
    q_{ab}=\sum_s q_s T^{(s)}_{ab},
    \qquad
    \delta K_{ab}=\sum_s k_s T^{(s)}_{ab},
\end{eqnarray}
the Lorentzian reality conditions require $q_s$ and $k_s$ to be real on the real metric section.

The full perturbation of the self-dual connection contains both the spin-connection perturbation and the extrinsic-curvature perturbation,
\begin{eqnarray}
    \delta A=\delta\Gamma[q]+i\delta K.
\end{eqnarray}
In TT gauge, the torsion-free condition gives
\begin{eqnarray}
{
    \delta\Gamma_s=+\frac{s}{2}q_s,
}
\end{eqnarray}
and hence
\begin{eqnarray}
{
    a_s\equiv \delta A_s=+\frac{s}{2}q_s+i k_s .
}
    \label{eq:tt_connection_perturbation}
\end{eqnarray}
This is the inhomogeneous analogue of the spin-connection correction included in the Bianchi~IX calculation above.  As we have seen, a real metric perturbation is not only an imaginary connection perturbation: it also changes the real part of the self-dual connection through $\delta\Gamma[q]$.

The quadratic expansion of the Chern--Simons functional about the de~Sitter connection takes the schematic form
\begin{eqnarray}
{
    S_{\rm CS}^{(2)}=-\frac12\sum_s \eta_s\,s\,a_s^2,
    \qquad
    \eta_s>0,
}
    \label{eq:csk_tt_quadratic}
\end{eqnarray}
where the positive constants $\eta_s$ depend on the normalization of the tensor harmonics.  Substituting Eq.~\eqref{eq:tt_connection_perturbation}, one finds
\begin{eqnarray}
{
    S_{\rm CS}^{(2)}
    =-\frac12\sum_s\eta_s\,s
    \left(\frac{s^2}{4}q_s^2-k_s^2+i s q_s k_s\right).
}
\end{eqnarray}
The real shear term on the Lorentzian real section is therefore
\begin{eqnarray}
{
    {\rm Re}\,S_{\rm CS}^{(2)}\supset
    +\frac12\sum_s\eta_s\,s\,k_s^2.
}
    \label{eq:tt_signed_shear}
\end{eqnarray}
{Thus, with this tensor-curl convention, the naive real contour gives Gaussian damping for $s<0$ but anti-damping for $s>0$.  Equivalently, the bare chiral CSK functional sees the signed operator $-{\cal C}$ rather than a positive helicity-even kernel.}  This is not a classical instability of de~Sitter: the ordinary ADM tensor Hamiltonian is positive for both helicities.  Rather, it is a statement about the contour-dependent probabilistic interpretation of the holomorphic CSK functional.

The Chern--Simons--Kodama functional is an exact holomorphic solution of the self-dual constraints in connection polarization, but the physical metric-space weight is defined only after imposing the Lorentzian reality conditions through a contour or inner product.  In minisuperspace the metric-polarization transform was sufficient to recover a positive Gaussian in the homogeneous anisotropies.  In the full TT sector, the naive real Lorentzian contour exposes the signed-curl obstruction in Eq.~\eqref{eq:tt_signed_shear}.  Therefore the question of tensor isotropization beyond minisuperspace is intrinsically a question about the contour completion of the CSK functional.

A natural completion is to choose the regular, no-boundary contour for the transverse graviton modes.  {Mode by mode this contour replaces the signed CSK tensor kernel by a positive helicity-even regularity kernel,}
\begin{eqnarray}
{
    -s\longrightarrow \Omega_s^{\rm NB},
    \qquad
    \Omega_s^{\rm NB}>0,
    \qquad
    \Omega_{n,+}^{\rm NB}=\Omega_{n,-}^{\rm NB}.
}
    \label{eq:abs_curl_replacement}
\end{eqnarray}
{For the round $S^3$ tensor modes used below, this kernel is proportional to $(n^2-1)/n$; at the level of signs this replacement is often summarized schematically as $-{\cal C}\rightarrow |{\cal C}|$.} {This choice also removes the leading quadratic tensor-helicity asymmetry of the connection-polarized CSK representative: the transverse kernel is helicity-even, $\Omega_{n,+}^{\rm NB}=\Omega_{n,-}^{\rm NB}$. Other admissible contour completions need not have this property.\footnote{The naive CSK contour is maximally chiral at the level of the quadratic tensor kernel: with the present tensor-curl convention, the $s<0$ sector is damped while the $s>0$ sector is anti-damped. It is therefore natural to ask whether a trade-off exists between the strength of isotropization and the amount of tensor-curl/chiral asymmetry retained by a contour-completed CSK functional.}}
The transverse wavefunctional then takes the form
\begin{eqnarray}
    \Psi_{\perp}^{\rm NB}[q]
    \simeq
    \exp\left[-\frac{1}{2\hbar}\sum_s\Omega_s q_s^2\right],
    \qquad
    \Omega_s>0,
    \label{eq:tt_nb_gaussian}
\end{eqnarray}
where $\Omega_s$ is the positive regularity kernel of the corresponding tensor mode.  Consequently},
{\begin{eqnarray}
    |\Psi_{\perp}^{\rm NB}[q]|^2
    \propto
    \exp\left[-\frac{1}{\hbar}\sum_s\Omega_s q_s^2\right].
    \label{eq:tt_nb_probability}
\end{eqnarray}
All transverse tensor perturbations are therefore probabilistically suppressed.  At quadratic order,} the {isotropic channel remains} the unique unstable/outgoing} channel{, while every TT departure from isotropy is confined by a positive Gaussian kernel.

The contour choice in Eq.~\eqref{eq:abs_curl_replacement} is a choice of integration cycle defining the metric-space state associated with the holomorphic CSK functional}.  In {particular,} the {no-boundary transverse contour is} the {same transverse contour that appears} in the {Hartle--Hawking state}.  {The distinction between the CSK functional considered here and the ordinary Hartle--Hawking wavefunctional is instead in the isotropic direction: the former is placed on an outgoing Gamow/Siegert contour, while the latter uses the no-boundary contour in the isotropic direction as well.

Once the transverse contour defining the physical state is specified, a robust quadratic isotropization mechanism can emerge beyond minisuperspace.  The simplest covariant contour realizing this result is the no-boundary transverse contour just described.  In Appendix~\ref{app:plebanski-completion} we state a minimal formal path-integral completion in which this contour arises naturally from a Plebanski construction with outgoing boundary conditions in the isotropic channel and regular no-boundary boundary conditions in the transverse graviton sector.  This completion is not meant to be unique.  Indeed, the results of Section~\ref{sec:confinement} suggest that a more hybrid perturbative contour, in which the reality-conditioned CSK contour is retained for the diagonal Bianchi~IX modes $(\beta_{\pm},p_{\pm})$ while a no-boundary contour is used for the remaining physical transverse-traceless modes, may also display the isotropization mechanism.  Such a hybrid choice is less obviously realized by a covariant nonperturbative bulk contour, however.  Appendix~\ref{app:plebanski-completion} therefore focuses on the cleaner Plebanski realization in which the CSK boundary no-boundary contour and the CSK boundary phase are formulated together to provide a robust, quantum gravitational mechanism for isotropization of de Sitter cosmologies.}

\section{Inflationary extension and robustness of the isotropization mechanism}

The analysis so far has been carried out for pure gravity with a positive cosmological constant{.  In} the {full homogeneous ADM variables the quadratic constraint} near the de~Sitter neck is
\begin{equation}
\begin{split}
  \mathcal H_{\rm grav}^{(2)}
  ={}& \frac{3}{4}a^4\left(-1+\frac{\Lambda}{3}a^2\right)
      -\frac{p_\alpha^2}{48}
      +\frac{p_+^2+p_-^2}{48} \\
      &+6a^4(\beta_+^2+\beta_-^2)+O(3),
\end{split}
\label{eq:Hgrav-minimal-integrated}
\end{equation}
where $(\beta_\pm,p_\pm)$ are the anisotropic {ADM variables introduced above}.  The {isotropic momentum $p_\alpha$ is} the {negative direction, while the anisotropy block} is {positive definite}.

To incorporate inflation in the minimal possible way, we add a single homogeneous inflaton $\phi$ with momentum $\pi_\phi$ and potential $V(\phi)$.  Its homogeneous Hamiltonian contribution is
\begin{equation}
  H_\phi = \frac{\pi_\phi^2}{2a^3} + a^3 V(\phi),
\end{equation}
{up to} the {same overall homogeneous volume normalization used in} the gravitational {constraint.  To leading order in slow roll,} the {potential shifts the cosmological term according to
\begin{equation}
  \Lambda_{\rm eff}(\phi)=\Lambda+8\pi G\,V(\phi),
\end{equation}
with the obvious specialization $\Lambda_{\rm eff}=8\pi G V$ if there is no separate bare cosmological constant.  The total homogeneous constraint is therefore captured}, to the order needed here, by
\begin{equation}
  \mathcal H_{\rm tot}^{(2)}
  =\mathcal H_{\rm grav}^{(2)}\big|_{\Lambda\rightarrow \Lambda_{\rm eff}(\phi)}
  +\frac{\pi_\phi^2}{2a^3}+O(\epsilon_V,\eta_V,3),
  \label{eq:Htot-integrated}
\end{equation}
where $O(3)$ denotes terms cubic in the homogeneous perturbations.

Assuming $V(\phi)$ drives slow-roll inflation, the parameters
$\epsilon_V = \frac{M_{\rm Pl}^2}{2}(V'/V)^2 \ll 1$ and
$\eta_V = M_{\rm Pl}^2(V''/V) \ll 1$ ensure that the kinetic term is subdominant and that the background satisfies
\begin{equation}
  3H\dot{\phi} \simeq -V'(\phi),
  \qquad
  H^2 \simeq \frac{\Lambda_{\rm eff}(\phi)}{3}.
\end{equation}
Here overdots denote derivatives with respect to the coordinate time $t$ in the gauge $N=1$. When we refer to ``evolution'' in this section we mean relational evolution after choosing a clock (conveniently $\phi$ on the slow-roll expanding branch), and one may rewrite derivatives via $d/dt=\dot\phi\,d/d\phi$ if desired.  The effective cosmological term varies only slowly,
\begin{equation}
  \frac{1}{\Lambda_{\rm eff}}\frac{d\Lambda_{\rm eff}}{dt}
  = \mathcal{O}(H \epsilon_V),
\end{equation}
so over one Hubble time its fractional change is of order $\epsilon_V \ll 1$.

In this regime the modified inflationary {Chern-Simons-Kodama} functional is {only needed} schematically, to leading adiabatic/slow-roll order{,
\[
\Psi^{(L)}_{\text{Kod}}[A,\phi]
=
\mathcal{N}
\exp\!\left[
\frac{3}{\ell_{\text{Pl}}^{2}\,\Lambda_{\rm eff}(\phi)}\,
S_{\text{CS}}[A,\phi]
\right].
\]
Exact inflationary solutions of this type were} found by the authors in isotropic and Bianchi-I {settings} \cite{ASM,Mal}.  We leave a full Bianchi{~}IX analysis using the same methods for future exploration.  The exact inflationary solutions modulate the {Chern--Simons} functional by the inflationary potential, providing a first order differential equation for that modulation function to be solved {self-consistently} with the deparametrized Hamiltonian constraint.

The sphaleron sector therefore evolves in an approximately de~Sitter background whose cosmological constant varies only adiabatically.  To leading order, the replacement $\Lambda\rightarrow\Lambda_{\rm eff}(\phi)$ in Eq.~\eqref{eq:Hgrav-minimal-integrated} captures the leading effect of the inflaton on the gravitational dynamics.

It is instructive to observe that, in the isotropic limit, the Hamiltonian constraint takes on a familiar form.  Using the standard minisuperspace relation {$\pi_\phi=a^3\dot\phi$}, the scalar kinetic term becomes $(1/2)a^3\dot\phi^{\,2}$.  Meanwhile the gravitational factor $-1+\Lambda_{\rm eff}(\phi)a^2/3$ is precisely the closed-FRW combination $a^2H^2-1$, so that the constraint reduces to the standard slow-roll Friedmann equation,
\begin{equation}
  H^2 + \frac{k}{a^2}
  \simeq
  \frac{\Lambda_{\rm eff}(\phi)}{3}
  + \frac{8\pi G}{3}\left(\frac{1}{2}\dot{\phi}^{\,2}\right).
\end{equation}
This shows explicitly that $\pi_\phi$ encodes the slow-roll kinetic energy and that $\phi$ may be used as a relational clock variable along the isotropic trajectory.  Once anisotropies are switched off, the full system reduces smoothly to ordinary slow-roll inflation in a $k=+1$ slicing.

The {anisotropy sector} is equally transparent{.  At fixed slowly varying $\Lambda_{\rm eff}$} the quadratic anisotropy block becomes
\begin{equation}
  \mathcal H_{\rm ani}^{(2)}
  =\frac{p_+^2+p_-^2}{48}
  +6a^4(\beta_+^2+\beta_-^2),
\end{equation}
which remains positive definite.  {Equivalently}, after choosing the slow-roll expanding branch as clock, the anisotropy variables obey a damped oscillator equation of the schematic form
\begin{equation}
  \ddot\beta_\pm+3H\dot\beta_\pm
  +\omega_\beta^2(a,\Lambda_{\rm eff})\beta_\pm
  =O(\epsilon_V,\eta_V,\beta^2),
  \qquad \omega_\beta^2>0.
  \label{eq:ani-simple}
\end{equation}
Thus the confining structure in the anisotropic directions is untouched by the slow variation of the inflaton, while Hubble friction damps the homogeneous shear along the slow-roll branch.  The unstable isotropic direction likewise persists.  The minisuperspace isotropization mechanism therefore remains operative throughout the slow-roll regime, up to adiabatic corrections of order $\epsilon_V$ and $\eta_V$.

Looking ahead, it will be important to move beyond the adiabatic
approximation and analyze the full coupled inflaton--gravity dynamics with
a general slow-roll potential.  In forthcoming work we plan to solve the
complete Hamiltonian system, including the homogeneous inflaton momentum
and the full non-linear gravitational degrees of freedom, in the spirit of
the analysis of the authors \cite{ASM,Mal}.
Our goal is to obtain a non-perturbative inflationary generalization of the
Kodama--sphaleron framework, to characterize the evolution of the
wavefunctional across the slow-roll epoch, and to determine how the
instantaneous Kodama-like states deform as the inflaton traverses a
generic potential $V(\phi)$.  This program will allow us to study the
transition between the early-time large-$\Lambda_{\rm eff}(\phi)$ phase and
the late-time small-$\Lambda_{\rm eff}$ vacuum in a fully dynamical setting,
and to explore whether the isotropization mechanism identified here extends
throughout the entire inflationary trajectory without approximation.

\section{Hyperbolic and Flat Cosmologies}

In this section we consider the Kodama functional and its semiclassical sector on a 
cosmological background whose spatial slices are compact hyperbolic manifolds, 
obtained as closed quotients of $H^{3}$. Such negatively curved geometries provide a natural arena in which to probe the interplay between Chern--Simons dynamics and global spatial topology, features that remain invisible in the standard $S^3$ or $\mathbb{R}^3$ settings.

Although compact hyperbolic manifolds do not admit global homogeneity and therefore fall outside the strict Bianchi classification, we will use a Bianchi~V minisuperspace truncation on the universal cover $H^3$ as a proxy for the local homogeneous sector relevant to the Chern--Simons functional. Indeed, there exists no genuinely inhomogeneous metric on the universal covering space $H^3$ with constant negative curvature: all local geometries are isometric to the homogeneous hyperbolic metric, and any inhomogeneity arises solely through discrete global identifications in the quotient $H^3/\Gamma$\footnote{This follows from the rigidity of hyperbolic geometry: any complete, simply connected Riemannian manifold of constant negative curvature is locally isometric to $H^3$, and compact quotients inherit inhomogeneity only through their discrete identifications rather than through local metric degrees of freedom.}. It is therefore consistent to formulate the minisuperspace reduction on $H^3$, with the understanding that this neglects the genuinely inhomogeneous/global modes associated with the quotient, and thus serves only as a local/homogeneous probe of the $k=-1$ sector. This approach isolates the homogeneous connection variables governing the Chern--Simons functional while remaining agnostic about detailed global topological data of the closed hyperbolic spatial slices.

The metric
takes the form

\begin{eqnarray}
    ds^2 \;=\; -N^2(t)\,dt^2 \;+\; a_i^2(t)\,(\omega^i)^2,
\end{eqnarray}
where $N(t)$ is the lapse, $a_i(t)$ $(i=1,2,3)$ are the three directional
scale factors, and the one-forms $\omega^i$ furnish a left-invariant coframe
on the spatial Bianchi group.
A standard choice of left-invariant one-forms for Bianchi~V is
\begin{eqnarray}
    \omega^1 &=& dx,\\
    \omega^2 &=& e^{x}\,dy,\\
    \omega^3 &=& e^{x}\,dz,
\end{eqnarray}
which satisfy the Maurer--Cartan equations
\begin{eqnarray}
    d\omega^1 &=& 0,\\
    d\omega^2 &=& \omega^1\wedge\omega^2,\\
    d\omega^3 &=& \omega^1\wedge\omega^3.
\end{eqnarray}
The tetrad for this metric is then
\begin{eqnarray}
   && e^0 \;=\; N\,dt, \qquad
    e^1 \;=\; a_1\,\omega^1, \qquad \\
    && e^2 \;=\; a_2\,\omega^2, \qquad
    e^3 \;=\; a_3\,\omega^3,
\end{eqnarray}
leading to the self-dual Ashtekar connection given by
\begin{eqnarray}
    A^1 &=& i\,b_1\,\omega^1,\\[4pt]
    A^2 &=& i\,b_2\,\omega^2 \;-\; \frac{a_3}{a_1}\,\omega^3,\\[4pt]
    A^3 &=& i\,b_3\,\omega^3+\frac{a_2}{a_1}\,\omega^2 \;.
\end{eqnarray}
The imaginary pieces encode the extrinsic curvature, while the real
off-diagonal terms reflect the intrinsic spatial curvature of the
Bianchi~V group.
The curvature of the Ashtekar connection is
\begin{eqnarray}
    F^i \;=\; dA^i + \frac{1}{2}\,\epsilon^i{}_{jk}\,A^j\wedge A^k.
\end{eqnarray}
Treating $a_i$ and their time derivatives as constant on each homogeneous
slice, and using $d\omega^1=0$, $d\omega^2=\omega^1\wedge\omega^2$,
$d\omega^3=\omega^1\wedge\omega^3$, one finds
\begin{widetext}
\begin{eqnarray}
    F^1 &=& i\,\dot b_1\,dt\wedge\omega^1
    \;+\;\left(\frac{a_2 a_3}{a_1^2} - b_2 b_3\right)\,\omega^2\wedge\omega^3,
    \\[6pt]
    F^2 &=& i\,\dot b_2\,dt\wedge\omega^2
    \;-\;N\left(\frac{b_3}{a_1} - \frac{a_3}{a_1^2}\,b_1\right)\,dt\wedge\omega^3
    \;+\; i\left(b_2 - \frac{a_2}{a_1}\,b_1\right)\,\omega^1\wedge\omega^2
    \;+\;\left(b_1 b_3 - \frac{a_3}{a_1}\right)\,\omega^1\wedge\omega^3,
    \\[6pt]
    F^3 &=& i\,\dot b_3\,dt\wedge\omega^3
    \;+N\left(\frac{b_2}{a_1} - \frac{a_2}{a_1^2}\,b_1\right)\,dt\wedge\omega^2
    \;+\; i\left(b_3 - \frac{a_3}{a_1}\,b_1\right)\,\omega^1\wedge\omega^3+\;\left(\frac{a_2}{a_1} - b_1 b_2\right)\,\omega^1\wedge\omega^2
    \;.
\end{eqnarray}
\end{widetext}
These expressions explicitly exhibit the mixture of kinetic (time-derivative)
and spatial-curvature contributions to the {self-dual $\mathfrak{su}(2)_{\mathbb C}$ curvature}.

In terms of the canonical pair $(b_i,\Pi_i)$, where 
\begin{eqnarray}
    \Pi_1 := a_2 a_3, \qquad
    \Pi_2 := a_1 a_3, \qquad
    \Pi_3 := a_1 a_2.
\end{eqnarray}
the Hamiltonian constraint can be written as
\begin{widetext}
    \begin{eqnarray}
    \mathcal{H}
    &=&
    -\big(
        \Pi_1\Pi_2\,b_1 b_2
        + \Pi_1\Pi_3\,b_1 b_3
        + \Pi_2\Pi_3\,b_2 b_3
    \big)
    + 3\,\Pi_1^2
    + \Lambda\,\Pi_1\Pi_2\Pi_3.
\end{eqnarray}
\end{widetext}
The saddle points of the Hamiltonian in the $b_i$ directions are
obtained from
\begin{eqnarray}
    \frac{\partial \mathcal{H}}{\partial b_i} = 0
    \quad\Rightarrow\quad
    b_1 = b_2 = b_3 = 0,
\end{eqnarray}
for non-degenerate $\Pi_i$. One can notice that, unlike the Bianchi IX model, the saddle point does not correspond to an isotropic brunch since there is no restriction on $\Pi_i$. In fact, imposing the constraint at this saddle gives
\begin{eqnarray}
    && \mathcal{H}\big|_{b_i=0}
    \;=\; 3\,\Pi_1^2 + \Lambda\,\Pi_1\Pi_2\Pi_3
    \;=\; 0 \\
    && \Rightarrow\quad
    \Pi_1\big(3\Pi_1 + \Lambda\,\Pi_2\Pi_3\big) = 0,
\end{eqnarray}
which for non-vanishing $\Pi_1$ yields the saddle-point relation
\begin{eqnarray}
    3\Pi_1 + \Lambda\,\Pi_2\Pi_3 = 0.
\end{eqnarray}
In terms of the scale factors, using $\Pi_1=a_2 a_3$, $\Pi_2=a_1 a_3$, $\Pi_3=a_1 a_2$,
this condition becomes
\begin{eqnarray}
    a_1^2 \;=\; -\frac{3}{\Lambda},
\end{eqnarray}
while $a_2$ and $a_3$ remain free. For $\Lambda>0$ this stationary point lies off the real scale-factor section (since it would require $a_1^2<0$), and therefore does not furnish the real isotropic de~Sitter sphaleron found in the closed ($S^3$) case. This does not preclude de~Sitter space in $k=0$ or $k=-1$ slicings; it only indicates that no analogous isotropic sphaleron saddle appears here in this minisuperspace analysis.

It should also be noted that, conducting a very similar analysis for a flat universe ($k=0$) realized by a Bianchi I geometry, leads to a Hamiltonian constraint of the form:
\begin{widetext}
    \begin{eqnarray}
    \mathcal{H}
    &=&
    -\big(
        \Pi_1\Pi_2\,b_1 b_2
        + \Pi_1\Pi_3\,b_1 b_3
        + \Pi_2\Pi_3\,b_2 b_3
    \big)
    + \Lambda\,\Pi_1\Pi_2\Pi_3,
\end{eqnarray}
\end{widetext}
for which the difference when compared to Bianchi V is the absence of the $\Pi_1^2$ term. For this case, {imposing the constraint at $b_i=0$ gives $\Lambda\Pi_1\Pi_2\Pi_3=0$}, {so} the {real solution} is {necessarily degenerate: at least one $\Pi_i$ vanishes.  In the isotropic subspace this collapses to} $a_1=a_2=a_3=0$. This reinforces the fact that the isotropization mechanism is a specific feature of the $S^3$ geometry.

\section{Future Directions}

In this work we have argued that the Lorentzian Chern--Simons--Kodama 
functional is most naturally understood as a metastable configuration above de~Sitter space --- a 
sphaleron configuration.  The dynamics provides a striking quantum--gravitational mechanism for the 
isotropization of a closed universe: perturbations about the saddle-point 
decay towards the de~Sitter attractor, in close analogy with familiar 
sphaleron transitions interpolating between topologically distinct vacua in 
Yang--Mills theory.  The appearance of this behavior in the Lorentzian Kodama 
wavefunctional invites a broader question: from what deeper state (or 
family of states) does the Kodama resonance emerge, and into what more complete 
description must it ultimately decay?

A related issue is the well-known non-normalizability of the Lorentzian Kodama 
state under large gauge transformations.  This failure of gauge invariance is 
precisely what one expects of a boundary functional exhibiting a large-gauge 
anomaly, rather than of a genuine physical state in a closed theory.  In 
quantum field theory, such anomalies are often resolved by anomaly inflow~\cite{HarveyTASI}: the anomalous variation of a 
lower-dimensional boundary theory is cancelled by the variation of a 
higher-dimensional bulk action.  Remarkably, the anomalous transformation of 
the Chern--Simons--Kodama functional mirrors exactly the structure of 
Chern--Simons boundary actions arising in topological phases of gauge theory 
and gravity.  This suggests that the Kodama wavefunctional may be most 
naturally interpreted as the boundary term of a higher-dimensional topological 
gravitational theory: for example, the 5D or 7D constructions appearing in 
topological M-theory, in which bulk inflow cancels the boundary anomaly and 
restores large-gauge invariance~\cite{DijkgraafVafa,HarveyTASI}.  Under this interpretation, 
the non-normalizability of the Kodama functional is not a pathology of 
four-dimensional quantum gravity, but a signal that the Kodama functional 
belongs to a larger, anomaly-free framework in one higher dimension.

There are already suggestive indications that such an embedding exists.  As noted 
by Dijkgraaf and Vafa (see p.~11 of Ref.~\cite{DijkgraafVafa}), the 
Plebanski formulation of gravity with a cosmological constant emerges naturally 
within their topological M-theory construction.  
In this context, the Kodama functional may be viewed as the four-dimensional
boundary wavefunctional associated with a specific flux or topological sector
of the underlying String/M-theory description, rather than as an autonomous
state of four-dimensional gravity.
Since the Kodama functional is the semiclassical solution of Plebanski gravity 
for $\Lambda>0$, this suggests that the Kodama functional may correspond to a 
distinguished vacuum or flux sector in the Dijkgraaf--Vafa framework. 

A complementary perspective on the metastability of de~Sitter space arises in
String/M--theory, where four--dimensional de~Sitter geometry has been argued to
appear not as a true vacuum but as a coherent (Glauber--Sudarshan) state built
over a supersymmetric background and supported by higher--dimensional degrees
of freedom \cite{BrahmaDasguptaTatar2020,BrahmaDasguptaTatar2021,BernardoDasguptaEtAl2021}.
In this picture, de~Sitter space is intrinsically time--dependent and
metastable, a viewpoint that closely parallels the interpretation advanced
here for the Lorentzian Chern--Simons--Kodama functional as suggestive of a Gamow--type
resonance rather than a normalizable ground state.  From this standpoint, the
Kodama wavefunctional may be viewed as a String/M--theoretic sector in which de~Sitter
metastability is a generic feature.

If a completion of the Kodama resonance into a proper Lorentzian state of 
quantum gravity can be found, then the topological significance of the 
Chern--Simons number may become clear.  At present, its Euclidean interpretation 
as a measure, and the close analogy with the Yang--Mills sphaleron, suggest an 
interpretation as a measure on the topological data of the theory.  After the 
reality conditions are imposed, the Euclidean Chern--Simons term acts as a 
measure on field configurations, including a contribution from torsion that 
vanishes on shell.  In an anomaly-inflow lift to M-theory, however, saddles of 
the full wavefunctional would generically include torsion.  We therefore 
expect that the genuine Chern--Simons phase of the completed state, after 
large-gauge anomaly cancellation, will admit an interpretation as a winding 
number in a higher-dimensional theory with torsion data, which usually involve flux configurations in String-M theory.

From an Euclidean perspective, this phase will, under Wick rotation, contribute 
to the measure on the space of spacetime geometries, and may provide insight 
into whether the isotropizing, spherical (as opposed to hyperbolic) Kodama 
resonance is, or is not, favored generically in the preparation of a 
fully nonperturbative and anomaly-free quantum state.  This question may be 
relevant to discussions of the apparent ``fine tuning'' of initial conditions 
required to produce the observed level of isotropy in the universe.

The condition $\partial \Sigma=0$ is not only required for functional
differentiability of $S_{\rm CS}[A]$, but it also makes precise the action of large {$SU(2)$ spin-lift gauge transformations}.  On a closed three-manifold
(such as $S^{3}$), gauge transformations fall into homotopy classes labeled by
$w(g)\in\pi_{3}(\mathrm{SU}(2))=\mathbb{Z}$, and the Chern--Simons functional
shifts by an integer,
\begin{equation}
S_{\rm CS}[A^{g}] = S_{\rm CS}[A] + 8\pi^{2} w(g).
\end{equation}
Consequently the Kodama functional transforms as
\begin{equation}
\Psi_{\rm CSK}[A^{g}]
= \exp\!\big(8\pi^{2}\kappa\, w(g)\big)\,\Psi_{\rm CSK}[A]
\end{equation}
with $\kappa=\frac{3}{\Lambda\ell_{\rm Pl}^2}.$
In Euclidean signature this factor can be a pure phase and is compatible with
$\theta$--sectors, but in the Lorentzian self--dual theory $\kappa$ is not
purely imaginary, so the large-gauge factor has non-unit modulus.  The state
therefore acquires an exponentially growing/decaying component along the
winding-number direction, which is the origin of the familiar statement that
the Lorentzian Kodama functional is non-normalizable under large gauge
transformations.  The topology $S^{3}$ does not remove this effect, rather it
renders it sharp by providing a clean integer classification of large-gauge copies.  This is precisely the behavior expected of an anomalous boundary
functional, and it motivates either a topological dressing (Soo’s compensating
$\mathcal{N}_{M}$ factors and $\theta$--sectors) or an anomaly-inflow
completion in one higher dimension, in which the non-unit large-gauge
variation of the Chern--Simons term is cancelled by a bulk contribution.

 {
}
As to the question of the ultimate fate of the Kodama resonance, the present 
article should be understood as a perturbative indicator.  The metastability 
of the resonance above de~Sitter space, together with the fact that the 
Euclidean Kodama functional includes de~Sitter as a nonperturbative saddle, raises 
the intriguing question of whether the perturbative Lorentzian resonance is a 
signal of a more general metastability of de~Sitter space in quantum gravity.  
It is plausible that once the decay begins it may proceed beyond the 
perturbative regime.  Whether this provides a natural mechanism for the end of 
inflation, or offers insight into the ultimate fate of de~Sitter space itself, 
depends on a proper embedding of the resonance into a fully Lorentzian and 
UV-complete quantum gravitational framework, such as may be realized in the 
M-theory scenario discussed above.

Altogether, these observations point towards a coherent but still unexplored 
picture: the Lorentzian Kodama functional may represent the semiclassical 
boundary manifestation of a higher-dimensional topological sector, whose bulk 
dynamics resolve both its gauge-theoretic anomaly and its metastable character.  
Understanding this completion, and the mechanism by which the Kodama resonance 
is excited in the early universe, promises to illuminate the role of 
self-dual gravitational configurations in cosmology and quantum gravity.  We 
leave these questions for future work.

Nonetheless, once the dynamics evolve towards the sphaleron saddle at the 
self-dual point, the structure of the Lorentzian Chern--Simons potential 
provides a natural mechanism for the isotropization of de~Sitter space.  In 
our analysis, the single unstable ``Kodama direction'' and the two stable 
anisotropic modes cooperate to funnel generic homogeneous perturbations toward 
the isotropic de~Sitter configuration.  This behavior offers a concrete 
realization of a quantum--gravitational isotropization mechanism: the decay of 
the Kodama resonance dynamically suppresses anisotropies and selects an 
approximately de~Sitter geometry.  It is tempting to speculate that this 
sphaleron-driven relaxation process may play a role in preparing the universe 
for an inflationary phase, or even serve as a seed for inflation itself.

It is useful to note that the dynamical picture described above admits a
precise analogy with the theory of quantum resonances, as originally
formulated by Gamow.  In ordinary quantum mechanics, unstable equilibria are
described by inverted harmonic oscillator Hamiltonians, whose outgoing WKB
solutions define Gamow states: non-normalizable wavefunctions on the real
configuration space that become normalizable only after analytic
continuation along an appropriate steepest-descent contour.  Such states are
characterized by complex energies and finite lifetimes, and they provide a
natural language for metastable decay processes.

In the Lorentzian Kodama framework analyzed here, the self-dual de~Sitter
configuration plays an analogous role.  Linearization of the minisuperspace
Hamiltonian about the sphaleron reveals an inverted quadratic potential along
the unique isotropic direction, while all anisotropic directions are
stabilized.  The corresponding Wheeler--DeWitt equation therefore admits
oscillatory WKB branches associated with outgoing evolution away from the
saddle.  When evaluated with the holomorphic inner product required by the
self-dual reality conditions, the integration contour in configuration space
is naturally deformed into the complex plane, converting the oscillatory
Lorentzian phase into an exponentially damped Gaussian.  In this sense, the
Lorentzian Kodama wavefunctional may be viewed as a Gamow-type resonance of
quantum geometry, with a characteristic decay rate (with respect to the chosen relational clock) set by
$\Omega \sim \sqrt{\Lambda/3}$. (Equivalently, $\Omega$ is of order the de~Sitter Hubble scale $H=\sqrt{\Lambda/3}$; in the closed slicing $a(t)=\sqrt{3/\Lambda}\cosh(Ht)$ this is the unique curvature scale of the saddle, so one expects a characteristic timescale $\tau\sim\Omega^{-1}$.) If one assumes a Gamow--type resonance interpretation for the Kodama functional, with complex energy $E=E_R-i\Gamma/2$ so that probability decays as $e^{-\Gamma T/\hbar}$ in the chosen relational time $T$, then the lifetime is $\tau\equiv\hbar/\Gamma$.  For an inverted-oscillator instability characterized by $\Omega$, one expects $\Gamma\sim\hbar\Omega$ (up to an order-one factor), and hence again $\tau\sim\Omega^{-1}$.

We emphasize, however, that this resonance interpretation is not required for
the isotropization mechanism identified in this work.  The suppression of
anisotropies and the dynamical selection of an isotropic de~Sitter geometry
follow directly from the sphaleron structure of the Hamiltonian and the large-gauge/framing topology encoded by the Chern--Simons functional.  The Gamow picture should therefore
be regarded as an interpretive layer that provides additional physical
intuition for the metastable character of the Lorentzian Kodama functional,
rather than as an independent assumption or organizing principle.
\section{Discussion and Outlook}

The analysis presented in this work points to a clear physical interpretation
of the Lorentzian Chern--Simons--Kodama construction.  The central result is not
the identification of a particular ground state, but the structure of the
quantum dynamics in the space of homogeneous geometries.  For a closed universe
with spherical spatial topology, these dynamics are organized around a
distinguished saddle-point configuration.  Near this saddle, the quantum
Hamiltonian exhibits a robust and physically transparent pattern: there is a
single unstable direction associated with isotropic expansion, while all
anisotropic degrees of freedom are stable and probabilistically disfavored, or dynamically suppressed.

This structure provides a concrete quantum-gravitational mechanism for
isotropization.  Generic homogeneous perturbations are driven towards the
isotropic subspace, not because anisotropies are assumed to be small, but
because they are disfavored by the quantum dynamics themselves.  The emergence
of an approximately de~Sitter geometry is therefore a dynamical outcome of the
quantum theory, rather than a consequence of fine-tuned initial conditions or
purely classical expansion effects.

The isotropization mechanism discussed here is fundamentally quantum mechanical in nature, while  The classical homogeneous geometry serves only to identify the relevant saddle-point configuration.  The physical consequences, such as stability and the suppression of anisotropy, arise only after solving the quantum constraint equations in the neighborhood of this saddle and imposing the appropriate inner product {and integration contour}.  {In} the reduced Bianchi~IX sector the Chern--Simons--Kodama wavefunctional is {Gaussian-suppressed} in the {homogeneous anisotropies after the Lorentzian reality conditions are implemented in metric polarization.  In the inhomogeneous transverse--traceless sector, the bare chiral CSK functional has} a {signed, chiral kernel, while a no-boundary transverse contour yields a positive Gaussian kernel for all tensor modes.  These two results together support} the isotropization mechanism identified here.

A central and nontrivial feature of the dynamics is that the isotropic mode is
the unique unstable direction.  In the full Bianchi~IX system, the quadratic
structure of the quantum Hamiltonian contains precisely one negative direction,
aligned with the isotropic combination of degrees of freedom, while the
remaining anisotropic modes are stabilized by a positive quantum potential.
As a result, quantum fluctuations that would otherwise generate shear or
directional structure are dynamically confined, whereas evolution along the
isotropic direction is allowed.  Isotropy is therefore not merely a late-time
attractor but a property selected directly by the quantum dynamics.

At the same time, the local structure of the quantum dynamics near the saddle
shares notable features with unstable configurations familiar from quantum
mechanics.  In particular, the Kodama wavefunctional exhibits behavior
characteristic of a Gamow-type state, with a natural frequency scale set by
the cosmological constant, $\Omega \sim \sqrt{\Lambda/3}$.  This scale governs
both the curvature of the quantum potential near the saddle and the width of
the semiclassical wavefunctional.  While this resonance-like behavior is not
the essential result of the present work, it provides a useful diagnostic of
the instability structure and suggests a deeper dynamical interpretation that
merits further investigation.

This quantum mechanism complements, and in important respects extends,
the classical picture of isotropization. Many WKB-like solutions of the
Wheeler--DeWitt equation can be decomposed, at least locally, into
semiclassical branches associated with distinct classical saddle
geometries. For branches that are homogeneous, expanding, and fall
within the hypotheses of the cosmic no-hair theorems, anisotropy is
expected to decay at late times under the classical evolution
\cite{Wald1983,MaleknejadSheikhJabbari2012}. The spatially closed
Bianchi~IX case studied here, however, lies outside the scope of
Wald's original result. Our mechanism therefore fills a natural gap:
it provides a short-time, intrinsically quantum-gravitational route to
isotropization near the closed de~Sitter saddle, where the
wavefunctional itself suppresses homogeneous shear. By contrast, in
the Bianchi~I and Bianchi~V sectors analyzed above, where late-time
classical isotropization may occur under the no-hair
hypotheses, we find no analogous early-time minisuperspace
sphaleron mechanism. The approach to de~Sitter space identified here
thus reflects a selection principle rooted in the quantum state,
rather than in classical expansion alone.

The role of spatial topology is essential in this construction.  The absence
of an analogous saddle-point structure in flat or hyperbolic cosmologies
highlights the special significance of closed spatial geometry.  In the
closed case, the topology allows for a nontrivial organization of the quantum
configuration space within the homogeneous minisuperspace truncation, leading to a distinguished isotropic saddle with the
properties described above.  Understanding how this topological selection
mechanism generalizes beyond homogeneous models remains an important open
question.

An important direction for future work is to place this picture on firmer
dynamical footing by establishing whether the saddle-point structure
identified here admits a precise formulation as a metastable quantum
configuration with a well-defined lifetime.  In recent work, Susskind and
Maltz have argued that de~Sitter space itself should be viewed not as a true
ground state, but as a metastable object with properties analogous to a
resonance in quantum mechanics.  It would be valuable to determine whether
the quantum dynamics of the Chern--Simons--Kodama construction provide a
concrete realization of this idea within a canonical quantum-gravitational
framework, and whether a decay rate or lifetime can be defined beyond the
perturbative regime.

Several further directions are suggested by these results.  {The most important} is to {promote} the {contour prescription used in the transverse tensor sector} to {a fully nonperturbative definition of} the {state}.  In {Sec.~\ref{sec:beyond-minisuperspace} we showed that the bare chiral CSK functional has a helicity obstruction in the physical TT sector}, {while a no-boundary transverse contour restores a positive tensor kernel at quadratic order.  A full treatment should derive} this {contour from a covariant completion of} the {wavefunctional, clarify its relation to} the self-dual reality conditions, {and determine whether any} parity-odd signatures {survive beyond the parity-symmetric no-boundary transverse Gaussian}.  Another important task is to clarify the embedding of the Chern--Simons--Kodama construction into a fully ultraviolet-complete framework, and to understand how its perturbative consistency is maintained at the nonperturbative level.  Finally, it will be valuable to explore the cosmological implications of this quantum-gravitational isotropization mechanism, including its possible role in preparing the universe for an inflationary phase.

Taken together, these results point towards the possibility of a more complete physical picture in which isotropy emerges dynamically from quantum gravity itself.  The Lorentzian
Chern--Simons--Kodama construction, when interpreted through the structure of
its quantum dynamics, encodes a mechanism that suppresses anisotropic
configurations and selects an approximately de~Sitter geometry.  This provides
a concrete example of how quantum-gravitational effects may play an active role
in shaping the large-scale structure of spacetime.

{\appendix

\section{A Minimal Completion}
\label{app:plebanski-completion}

We have thus far used a transverse no-boundary contour to define the probability weight associated with the CSK functional beyond minisuperspace.  The purpose of this appendix is to spell out a minimal formal setting in which that contour may arise.  The discussion is not meant to provide a unique completion of the CSK functional, nor a rigorous nonperturbative definition of quantum gravity.  Rather, it gives a concrete path-integral object on the same formal footing as the Hartle--Hawking construction, while retaining the CSK boundary phase and the resonance interpretation of the isotropic direction.  Related recent work on de~Sitter quantum gravity has emphasized both the role of Hartle--Hawking-type sums in low-dimensional models and the subtleties of the no-boundary norm~\cite{CotlerJensenMaloney2020,CotlerJensen2026}.

It is useful to recall first that the relation between the CSK functional and the Hartle--Hawking wavefunctional is not accidental.  In isotropic minisuperspace, the Chern--Simons state is Fourier-dual to the Hartle--Hawking and Vilenkin wavefunctions, with the choice of integration contour in the connection variable selecting the corresponding branch~\cite{Magueijo:2020ugp}.  More generally, because the CSK functional solves the Hamiltonian constraint in connection polarization, its transform to triad or metric variables gives a formal generalized Hartle--Hawking wavefunctional; the explicit evaluation of this transform depends on the symmetry reduction and on the contour chosen~\cite{AlexanderHerczegMagueijo2021}.  Thus the same holomorphic CSK functional can underlie several different quantum-cosmological objects.  The functional itself is not yet a complete physical state until a reality-condition measure and an integration cycle have been specified.

The ordinary Hartle--Hawking choice is one such completion.  In the no-boundary state, the contour is chosen so that the four-dimensional geometry is regular in the interior and induces the prescribed real three-geometry on the boundary~\cite{HartleHawking}.  In Bianchi IX minisuperspace this no-boundary prescription is known to damp large anisotropic perturbations, and in an exactly solvable biaxial model it yields normalizable probabilities with low amplitude for large anisotropies~\cite{Dorronsoro2018,JanssenHalliwellHertog2019}.  This known no-boundary isotropization is consistent with the thesis of the present work: the CSK functional, once completed by an appropriate contour, gives rise to a probability weight that suppresses anisotropy.  The contour adopted in the body of this paper is not, however, the full Hartle--Hawking contour.  We instead regard the isotropic direction as a Gamow/Siegert direction of a CSK resonance, while assigning the stable transverse tensor modes the same regular no-boundary contour that appears in the Hartle--Hawking state.

The minimal path-integral realization of this idea is naturally written in self-dual Plebanski variables.  Let $E$ denote real boundary triad data on $\Sigma$, obtained from the pullback of the Plebanski two-form by
\begin{eqnarray}
    E^a_i=-\frac12\epsilon^{abc}\Sigma_{i\,bc}\big|_{\Sigma},
    \label{eq:appendix_boundary_triad_from_sigma}
\end{eqnarray}
up to the orientation convention on the boundary.  The Lorentzian reality conditions on the boundary are
\begin{eqnarray}
    E=\bar E,
    \qquad
    A+\bar A=2\Gamma(E).
    \label{eq:appendix_reality_conditions}
\end{eqnarray}
The bulk fields may be complex along the integration cycle, just as in the usual complex-contour formulation of the no-boundary path integral, but the boundary data are real.  A schematic form of the self-dual Plebanski action, in conventions compatible with Eq.~\eqref{eqSDaction}, is
\begin{eqnarray}
S_{\rm Pl}[A,\Sigma,{\Phi}]
&=&{i\over 8\pi G}\int_M
\left[
\Sigma_i\wedge F^i[A]
-{\Lambda\over 6}\Sigma_i\wedge\Sigma^i\right.\nonumber\\
&&\left.
+{\Phi_{ij}}\Sigma^i\wedge\Sigma^j
\right],
\label{eq:appendix_plebanski_action}
\end{eqnarray}
where $\Phi_{ij}$ schematically denotes the symmetric tracefree multiplier imposing the simplicity constraints.  Varying $\Phi_{ij}$ gives the tracefree condition
\begin{eqnarray}
    \Sigma^i\wedge\Sigma^j
    -\frac13\delta^{ij}\Sigma^k\wedge\Sigma_k=0,
    \label{eq:appendix_simplicity_constraint}
\end{eqnarray}
with the trace part separated into the cosmological term.  On the nondegenerate sector these constraints imply that the $\Sigma^i$ are the self-dual two-forms of a tetrad.

We define the CSK resonance with no-boundary transverse contour by
\begin{eqnarray}
\Psi_{\rm CSK}^{\rm res.}[E]
&=&
\int_{\substack{{\cal J}_{\rm out}^{(Q)}\times {\cal J}_{\rm NB}^{(\perp)}\\ E_\partial[\Sigma]=E}}
{\cal D}A\,{\cal D}\Sigma\,{\cal D}{\Phi} \nonumber\\
&&\times
\exp\left[{i\over\hbar}S_{\rm Pl}[A,\Sigma,{\Phi}]\right].
\label{eq:appendix_kodama_res_path_integral}
\end{eqnarray}
Here ${\cal J}_{\rm out}^{(Q)}$ denotes the outgoing contour in the isotropic throat variable $Q$, while ${\cal J}_{\rm NB}^{(\perp)}$ denotes the regular no-boundary contour in the transverse tensor directions.  By contrast, the ordinary Hartle--Hawking contour in the same variables would be
\begin{eqnarray}
{\cal J}_{\rm HH}
=
{\cal J}_{\rm NB}^{(Q)}\times {\cal J}_{\rm NB}^{(\perp)}.
\label{eq:appendix_hh_contour}
\end{eqnarray}
Thus the two prescriptions share the same transverse contour but differ in the isotropic direction.  The Hartle--Hawking state is regular/no-boundary in all directions; the Chern-Simons-Kodama resonance is outgoing in the isotropic direction and no-boundary only in the transverse directions.

It is useful to separate the connection-polarized amplitude from the final triad-polarized wavefunctional.  Define
\begin{eqnarray}
Z_{\cal J}[A_\partial]
&=&
\int_{\substack{{\cal J},\; A|_\Sigma=A_\partial}}
{\cal D}A\,{\cal D}\Sigma\,{\cal D}\Phi\;\nonumber\\
&&\times
\exp\left[{i\over\hbar}S_{\rm Pl}[A,\Sigma,\Phi]\right].
\label{eq:appendix_connection_polarized_Z}
\end{eqnarray}
Its semiclassical expansion has the form
\begin{eqnarray}
Z_{\cal J}[A_\partial]
\simeq
\sum_{\gamma\in\mathrm{saddles}({\cal J})}
 n_\gamma\,\Delta_\gamma[A_\partial]
\exp\left[{i\over\hbar}S_{\rm Pl}[\gamma]\right],
\label{eq:appendix_saddle_expansion_Z}
\end{eqnarray}
where $\Delta_\gamma$ is the one-loop determinant and $n_\gamma$ is the thimble coefficient.  The CSK functional appears as the leading contribution of the self-dual de~Sitter saddle.  On this saddle,
\begin{eqnarray}
F^i=\frac{\Lambda}{3}\Sigma^i,
\label{eq:appendix_self_dual_saddle}
\end{eqnarray}
so that $\Sigma^i=3F^i/\Lambda$.  Substituting into Eq.~\eqref{eq:appendix_plebanski_action} gives
\begin{eqnarray}
S_{{\rm Pl},{\rm on\ shell}}
=
{i\over 8\pi G}{3\over 2\Lambda}\int_M F^i\wedge F_i .
\label{eq:appendix_on_shell_csk}
\end{eqnarray}
With ${\operatorname{Tr}}(T_iT_j)=-\delta_{ij}/2$,
\begin{eqnarray}
dS_{\rm CS}={\operatorname{Tr}}(F\wedge F)
=-\frac12F^i\wedge F_i,
\end{eqnarray}
and hence, for a single boundary component with the orientation chosen to match the CSK convention,
\begin{eqnarray}
\int_M F^i\wedge F_i=-2S_{\rm CS}[A_\partial].
\end{eqnarray}
Therefore, using $\ell_{\rm Pl}^2=8\pi G\hbar$,
\begin{eqnarray}
\exp\left[{i\over\hbar}S_{{\rm Pl},{\rm on\ shell}}\right]
=
\exp\left[\frac{3}{\Lambda\ell_{\rm Pl}^2}S_{\rm CS}[A_\partial]\right].
\label{eq:appendix_csk_saddle_weight}
\end{eqnarray}
Thus, near this saddle,
\begin{eqnarray}
Z_{\cal J}[A_\partial]
&\simeq&
\Delta_{\cal J}[A_\partial]
\exp\left[\frac{3}{\Lambda\ell_{\rm Pl}^2}S_{\rm CS}[A_\partial]\right]
\nonumber\\
&&\times
\left(1+O(\hbar)\right),
\label{eq:appendix_Z_contains_CSK}
\end{eqnarray}
which makes explicit the sense in which the Plebanski saddle expansion begins with the CSK boundary functional.  The contour completion in Eq.~\eqref{eq:appendix_kodama_res_path_integral} should therefore be understood as a gravitational path integral whose connection-polarized saddle representative is the CSK functional, rather than as a hand modification of the bare connection-space state.

We now derive the transverse contour used in Sec.~\ref{sec:beyond-minisuperspace}.  Around the round de~Sitter neck, expand the real boundary metric as
\begin{eqnarray}
    h_{ab}=a_\star^2(\gamma_{ab}+q_{ab}),
    \qquad
    K_{ab}=\delta K_{ab},\nonumber\\
    a_\star=H^{-1},
    \qquad
    H^2={\Lambda\over 3}.
\end{eqnarray}
After solving the linearized constraints, the physical inhomogeneous pure-gravity perturbations are transverse--traceless tensors.  Choose tensor harmonics satisfying
\begin{eqnarray}
D^a T^{(n,\lambda)}_{ab}=0,
\qquad
\gamma^{ab}T^{(n,\lambda)}_{ab}=0,\nonumber\\
{\cal C}T^{(n,\lambda)}_{ab}=\lambda n T^{(n,\lambda)}_{ab},
\label{eq:appendix_tensor_harmonics}
\end{eqnarray}
where {${\cal C}$ is the tensor curl on the unit three-sphere with the sign convention of Sec.~\ref{sec:beyond-minisuperspace}} and $\lambda=\pm$.  For a mode with curl eigenvalue $s=\lambda n$, the torsion-free spin-connection perturbation is
\begin{eqnarray}
{
    \delta\Gamma_s=+\frac{s}{2}q_s,
}
\end{eqnarray}
so that the self-dual connection perturbation is
\begin{eqnarray}
{
    a_s\equiv \delta A_s=+\frac{s}{2}q_s+i k_s.
}
    \label{eq:appendix_connection_perturbation}
\end{eqnarray}
The bare CSK quadratic functional is proportional to
\begin{eqnarray}
{
    S_{\rm CS}^{(2)}\sim -\frac12\sum_s \eta_s s a_s^2,
    \qquad
    \eta_s>0.
}
    \label{eq:appendix_signed_csk_kernel}
\end{eqnarray}
On the real Lorentzian section $q_s,k_s\in {\mathbb R}$, its real shear contribution contains
\begin{eqnarray}
{
    {\rm Re}\,S_{\rm CS}^{(2)}
    \supset
    +\frac12\sum_s \eta_s s k_s^2.
}
\end{eqnarray}
{This is the signed kernel discussed in the body of the paper.  With the present tensor-curl convention, the exponent is Gaussian-damped for $s<0$ and Gaussian-enhanced for $s>0$ on the naive real contour.  Thus the $s<0$ sector is the damped tensor-curl sector and the $s>0$ sector is the anti-damped tensor-curl sector.  This assignment of signs is only a relabeling of the two curl sectors relative to the opposite convention.}

The Plebanski no-boundary contour replaces this signed boundary kernel by the regular bulk Dirichlet-to-Neumann map.  To see this explicitly, consider the Euclidean de~Sitter cap
\begin{eqnarray}
    ds_E^2=H^{-2}\left(d\chi^2+\sin^2\chi\,d\Omega_3^2\right),\nonumber\\
    0\leq \chi\leq \frac{\pi}{2}.
\end{eqnarray}
For a TT tensor mode the quadratic Euclidean action has the form
\begin{eqnarray}
I_E^{(2)}
&=&\frac{{\cal N}_T}{2}\sum_{n,\lambda}
\int_0^{\pi/2}d\chi\,\sin^3\chi\nonumber\\
&&\times
\left[
(q'_{n\lambda})^2+\frac{n^2-1}{\sin^2\chi}q_{n\lambda}^2
\right],
\label{eq:appendix_tt_euclidean_action}
\end{eqnarray}
with ${\cal N}_T>0$.  The mode equation is
\begin{eqnarray}
q''_{n\lambda}+3\cot\chi\,q'_{n\lambda}
-\frac{n^2-1}{\sin^2\chi}q_{n\lambda}=0.
\label{eq:appendix_tt_mode_equation}
\end{eqnarray}
It depends on $n$ but not on $\lambda$.  The regular solution at the south pole, normalized by $q_{n\lambda}(\pi/2)=q_{n\lambda}^{(0)}$, has logarithmic derivative
\begin{eqnarray}
\left.\frac{q'_{n\lambda}}{q_{n\lambda}}\right|_{\chi=\pi/2}
=\frac{n^2-1}{n}.
\label{eq:appendix_dirichlet_neumann}
\end{eqnarray}
This result is positive and independent of helicity.  On shell, Eq.~\eqref{eq:appendix_tt_euclidean_action} reduces to the boundary term
\begin{eqnarray}
I_{E,{\rm on\ shell}}^{(2)}
=\frac{{\cal N}_T}{2}\sum_{n,\lambda}
\frac{n^2-1}{n}\left(q_{n\lambda}^{(0)}\right)^2.
\end{eqnarray}
Defining
\begin{eqnarray}
    \Omega_n={\cal N}_T\frac{n^2-1}{n},
    \qquad
    \Omega_n>0,
\end{eqnarray}
one obtains
\begin{eqnarray}
\Psi_{\perp}^{\rm NB}[q]
\simeq
\exp\left[-\frac{1}{2\hbar}\sum_{n,\lambda}\Omega_n q_{n\lambda}^2\right].
\label{eq:appendix_transverse_nb_wavefunction}
\end{eqnarray}
Thus the regular no-boundary contour replaces the signed tensor-curl kernel by a positive helicity-even Dirichlet-to-Neumann kernel.  For the round $S^3$ modes this kernel is
\begin{eqnarray}
    \Omega_n={\cal N}_T\frac{n^2-1}{n},
    \qquad \Omega_{n,+}=\Omega_{n,-}>0,
\end{eqnarray}
{and at the level of signs this is the schematic replacement $-{\cal C}\rightarrow |{\cal C}|$.}  The probability weight becomes
\begin{eqnarray}
    |\Psi_{\perp}^{\rm NB}[q]|^2
    \propto
    \exp\left[-\frac{1}{\hbar}\sum_{n,\lambda}\Omega_n q_{n\lambda}^2\right].
\label{eq:appendix_transverse_probability}
\end{eqnarray}
All transverse tensor modes are therefore probabilistically suppressed.  This is the contour prescription used in Sec.~\ref{sec:beyond-minisuperspace}.

The reality conditions are implemented in this construction in the same way as in the Hartle--Hawking path integral.  The boundary triad is real and satisfies Eq.~\eqref{eq:appendix_reality_conditions}.  The off-shell integration cycle in the bulk is complex, and the extrinsic-curvature variable need not remain real along the contour.   In the present case, the complex transverse thimbles are selected by regularity of the Plebanski bulk solution, while the isotropic contour is selected by the outgoing resonance boundary condition.

Formally, Eq.~\eqref{eq:appendix_kodama_res_path_integral} satisfies the Hamiltonian and momentum constraints in the same sense as the Hartle--Hawking wavefunctional.  If the action, measure, and contour are invariant under the gauge transformations generated by the constraints, a deformation of the final boundary is a change of variables in the path integral.  Therefore
\begin{eqnarray}
    \widehat{\cal H}\Psi_{\rm CSK}^{\rm res.}[E]
    =
    {\cal B}_{\cal J}[E],
\end{eqnarray}
where ${\cal B}_{\cal J}$ is a possible boundary contribution from the ends of the contour.  For an admissible Lefschetz contour, this contribution vanishes, giving
\begin{eqnarray}
    \widehat{\cal H}\Psi_{\rm CSK}^{\rm res.}&=&0,
    \qquad
    \widehat{\cal H}_a\Psi_{\rm CSK}^{\rm res.}=0,\nonumber\\
    \widehat{\cal G}_i\Psi_{\rm CSK}^{\rm res.}&=&0.
\end{eqnarray}
This argument is formal, as it is for the usual Hartle--Hawking state.  We do not claim here to have constructed the full nonperturbative contour.  The point is that Eq.~\eqref{eq:appendix_kodama_res_path_integral} gives a minimal candidate completion in which the CSK boundary phase, the outgoing isotropic resonance, and the no-boundary transverse contour are all part of a single gravitational path integral.

The construction is compatible with effective field theory in the usual semiclassical regime.  If $M_\ast$ denotes the scale suppressing higher-curvature corrections, the saddle expansion is controlled when
\begin{eqnarray}
    H\ll M_{\rm Pl},
    \qquad
    H\ll M_\ast,
    \qquad
    \frac{n}{a_\star}\ll M_\ast
\end{eqnarray}
for the tensor modes retained in the effective description.  Local higher-curvature terms shift the regularity kernel by controlled corrections,
\begin{eqnarray}
    \frac{\delta\Omega_n}{\Omega_n}
    =
    {\cal O}\left(\frac{H^2}{M_\ast^2},\frac{n^2/a_\star^2}{M_\ast^2}\right),
\end{eqnarray}
and therefore cannot change the sign of the transverse Gaussian in the controlled regime.

Finally, this completion is not unique.  {The CSK functional is best viewed as a holomorphic seed in a complexified generalization of the Hartle--Hawking construction: different constraint-compatible integration cycles of the original gravitational path integral define different physical completions.  Choosing the no-boundary contour in both the isotropic and transverse directions gives the Hartle--Hawking state written in self-dual/Plebanski variables~\cite{HartleHawking,Dorronsoro2018,JanssenHalliwellHertog2019}.  Choosing an outgoing/tunneling contour in the isotropic direction while assigning the perturbations stable vacuum, Robin, or no-boundary boundary conditions gives a Vilenkin-like or resonance-type completion~\cite{Vilenkin1984,VilenkinYamada2018,DiTucciLehners2019}.  More generally, admissible Lorentzian Picard--Lefschetz thimbles of the original gravitational action also solve the original constraints, although only those whose induced TT kernel has positive real part perturbatively isotropize~\cite{FeldbruggeLehnersTurok2017}. }
Other completions may arise from anomaly-inflow or determinant-line structures associated with the CSK large-gauge transformation law~\cite{HarveyTASI,APS}, or from higher-dimensional topological constructions in which Plebanski gravity appears naturally, such as the topological M-theory framework of Dijkgraaf and Vafa~\cite{DijkgraafVafa}.  String/M-theoretic descriptions of de~Sitter space as a metastable coherent state provide another possible setting in which the resonance interpretation could be embedded~\cite{BrahmaDasguptaTatar2021,BrahmaDasguptaTatar2020,BernardoDasguptaEtAl2021}.  The role of the present appendix is only to exhibit the minimal completion required for the contour prescription used in this article.  It shows that the transverse no-boundary contour can be realized as the natural regularity contour of a Plebanski gravitational path integral---dubbed the Chern--Simons--Kodama Resonance---whose saddle contribution is the CSK boundary functional.}
\pagebreak
\section*{Acknowledgments}
We thank Heliudson Bernardo, Robert Brandenberger, Jordan Cotler, Keshav Dasgupta, Jo\~{a}o Magueijo, Evan McDonough, Laura Mersini-Houghton, Sylvia Viera, and Robert Wald for insightful discussions. We thank the late Joe Polchinski who many years ago inspired the use of Sphalerons in de Sitter space, G.~W.~Gibbons and A.~R.~Steif for their Lorentzian construction that inspired this work, and colleagues in the Brown Center for Theoretical Physics and Innovation for discussions on self-dual gravity and Chern--Simons theory.

D.L.D. acknowledges support as a Black Hole Initiative Fellow. His contribution to this publication is funded in part by the Gordon and Betty Moore Foundation (Grant \#13526). It was also made possible through the support of a grant from the John Templeton Foundation (Grant \#63445). The opinions expressed in this publication are those of the author(s) and do not necessarily reflect the views of these Foundations. The Flatiron Institute is supported by the Simons Foundation.


\balance   
\end{document}